\documentclass[twocolumn]{bmcart}

\pdfoutput=1

\usepackage{amsthm,amsmath}
\usepackage{array}
\usepackage{diagbox}
\usepackage[utf8]{inputenc} 
\usepackage{textcomp}
\usepackage{multirow}

\usepackage{graphicx}

\usepackage[single]{acro}
\usepackage[capitalize]{cleveref}
\crefname{table}{Tab.}{Tabs.}
\Crefname{table}{Tab.}{Tabs.}

\DeclareAcronym{VLLS}{
	short=Vienna \acs{5G} \ac{LL} Simulator,
	long=Vienna \acs{5G} \ac{LL} Simulator,
	class=nolist
}
\DeclareAcronym{3GPP}{
	short=3GPP,
	long=3rd Generation Partnership Project
}
\DeclareAcronym{4G}{
	short=4G,
	long=fourth generation,
	class=nolist
}
\DeclareAcronym{5G}{
	short=5G,
	long=fifth generation,
	class=nolist
}
\DeclareAcronym{AMC}{
	short=AMC,
	long=adaptive modulation and coding
}
\DeclareAcronym{NR}{
	short=NR,
	long=new radio
}
\DeclareAcronym{BLER}{
	short=BLER,
	long=block error ratio
}
\DeclareAcronym{CRAN}{
	short=C-RAN,
	long=cloud \ac{RAN}
}
\DeclareAcronym{CDL}{
	short=CDL,
	long=clustered delay line
}
\DeclareAcronym{QAM}{
	short=QAM,
	long=quadrature amplitude modulation
}
\DeclareAcronym{CPOFDM}{
	short=CP-OFDM,
	long=cyclic prefix \ac{OFDM}
}
\DeclareAcronym{CSI}{
	short=CSI,
	long=channel state information
}
\DeclareAcronym{DFT}{
	short=DFT,
	long=discrete Fourier transform
}
\DeclareAcronym{DFTSOFDM}{
	short=DFT-s-OFDM,
	long=\ac{DFT} spread \ac{OFDM}
}
\DeclareAcronym{FBMC}{
	short=FBMC,
	long=filter-bank multicarrier
}
\DeclareAcronym{FDD}{
	short=FDD,
	long=frequency-division duplexing
}
\DeclareAcronym{FOFDM}{
	short=f-OFDM,
	long=filtered \ac{OFDM}
}
\DeclareAcronym{FOM}{
	short=FOM,
	long=flexible/open/modular
}
\DeclareAcronym{GFDM}{
	short=GFDM,
	long=generalized frequency division multiplexing
}
\DeclareAcronym{LDPC}{
	short=LDPC,
	long=low-density parity-check
}
\DeclareAcronym{LL}{
	short=LL,
	long=link level
}
\DeclareAcronym{LTE}{
	short=LTE,
	long=Long Term Evolution
}
\DeclareAcronym{LTEA}{
	short=LTE-A,
	long=\ac{LTE}-Advanced
}
\DeclareAcronym{MIMO}{
	short=MIMO,
	long=multiple-input multiple-output
}
\DeclareAcronym{ML}{
	short=ML,
	long=maximum likelihood
}
\DeclareAcronym{MMSE}{
	short=MMSE,
	long=minimum mean squared error
}
\DeclareAcronym{NOMA}{
	short=NOMA,
	long=non-orthogonal multiple access
}
\DeclareAcronym{OMA}{
	short=OMA,
	long=orthogonal multiple access
}
\DeclareAcronym{PHY}{
	short=PHY,
	long=physical layer
}
\DeclareAcronym{OFDM}{
	short=OFDM,
	long=orthogonal frequency division multiplexing
}
\DeclareAcronym{RAN}{
	short=RAN,
	long=radio access network
}
\DeclareAcronym{SDN}{
	short=SDN,
	long=software defined networking
}
\DeclareAcronym{SL}{
	short=SL,
	long=system level
}
\DeclareAcronym{SNR}{
	short=SNR,
	long=signal-to-noise ratio
}
\DeclareAcronym{TDD}{
	short=TDD,
	long=time-division duplexing
}
\DeclareAcronym{TDL}{
	short=TDL,
	long=tapped delay line
}
\DeclareAcronym{UFMC}{
	short=UFMC,
	long=universal filtered multicarrier
}
\DeclareAcronym{UMTS}{
	short=UMTS,
	long=universal mobile telecommunications system
}
\DeclareAcronym{VCCS}{
	short=VCCS,
	long=Vienna Cellular Communications Simulators
}
\DeclareAcronym{WIMAX}{
	short=WiMAX,
	long=worldwide interoperability for microwave access
}
\DeclareAcronym{WOLA}{
	short=WOLA,
	long=weighted overlap and add
}
\DeclareAcronym{BS}{
	short=BS,
	long=base station
}
\DeclareAcronym{UE}{
	short=user,
	long=user,
	class=nolist
}
\DeclareAcronym{FER}{
	short=FER,
	long=frame error ratio
}
\DeclareAcronym{BER}{
	short=BER,
	long=bit error ratio
}
\DeclareAcronym{CRC}{
	short=CRC,
	long=cyclic redundancy check
}
\DeclareAcronym{BCJR}{
	short=BCJR,
	long=Bahl-Cocke-Jelinek-Raviv
}
\DeclareAcronym{SC}{
	short=SC,
	long=successive cancellation
}
\DeclareAcronym{SIC}{
	short=SIC,
	long=successive interfernce cancellation
}
\DeclareAcronym{MUST}{
	short=MUST,
	long=multi-user superposition transmission
}
\DeclareAcronym{PMI}{
	short=PMI,
	long=precoding matrix indicator
}
\DeclareAcronym{RI}{
	short=RI,
	long=rank indicator
}
\DeclareAcronym{CQI}{
	short=CQI,
	long=channel quality indicator
}
\DeclareAcronym{OLSM}{
	short=OLSM,
	long=open loop spatial multiplexing
}
\DeclareAcronym{CLSM}{
	short=CLSM,
	long=closed loop spatial multiplexing
}

\DeclareAcronym{AWGN}{
	short=AWGN,
	long=additive white Gaussian noise
}
\DeclareAcronym{SINR}{
	short=SINR,
	long=signal to interference and noise ratio
}
\DeclareAcronym{ZF}{
	short=ZF,
	long=zero forcing
}
\DeclareAcronym{CP}{
	short=CP,
	long=cyclic prefix
}
\DeclareAcronym{ISI}{
	short=ISI,
	long=intersymbol interference
}
\DeclareAcronym{ICI}{
	short=ICI,
	long=intercarrier interference
}
\DeclareAcronym{FFT}{
	short=FFT,
	long=fast Fourier transform
}
\DeclareAcronym{OOB}{
	short=OOB,
	long=out of band 
}
\DeclareAcronym{IFFT}{
	short=IFFT,
	long=inverse \acl{FFT}
}
\DeclareAcronym{ZP}{
	short=ZP,
	long=zero prefix
}
\DeclareAcronym{EMBB}{
	short=eMBB,
	long=enhanced mobile broadband
}
\DeclareAcronym{MMTC}{
	short=mMTC,
	long=massive machine-type communication
}
\DeclareAcronym{URLLC}{
	short=uRLLC,
	long=ultra-reliable and low-latency communication
}
\DeclareAcronym{MCS}{
	short=MCS,
	long=modulation and coding scheme
}
\DeclareAcronym{IUI}{
	short=IUI,
	long=inter-user interference
}
\DeclareAcronym{RMS}{
	short=rms,
	long=root mean squared
}
\DeclareAcronym{MMW}{
	short=mmWave,
	long=millimeter wave
}
\DeclareAcronym{FEC}{
	short=FEC,
	long=forward error correction
}
\DeclareAcronym{PDSCH}{
	short=PDSCH,
	long=physical downlink shared channel
}
\DeclareAcronym{PUSCH}{
	short=PUSCH,
	long=physical uplink shared channel
}
\DeclareAcronym{TWDP}{
	short=TWDP,
	long=two-wave with diffuse power
}

\startlocaldefs
\newcolumntype{P}[1]{>{\centering\arraybackslash}p{#1}}
\newcommand{\matlab}{\textsc{Matlab}}
\renewcommand{\arraystretch}{1.2}
\endlocaldefs

\begin{document}

\begin{frontmatter}

\begin{fmbox}
\dochead{Research}


\title{Versatile Mobile Communications Simulation:\\The Vienna 5G Link Level Simulator}


\author[
   addressref={aff1,aff2},              
   corref={aff1},                       
   email={spratsch@nt.tuwien.ac.at}   
]{\inits{SP}\fnm{Stefan} \snm{Pratschner}}
\author[
	addressref={aff1,aff2},
	email={btahir@nt.tuwien.ac.at}
]{\inits{BT}\fnm{Bashar} \snm{Tahir}}
\author[
	addressref={aff1,aff2},
	email={lmarijan@nt.tuwien.ac.at}
]{\inits{LM}\fnm{Ljiljana} \snm{Marijanovic}}
\author[
	addressref={aff2},
	email={mmussbah@nt.tuwien.ac.at}
]{\inits{MM}\fnm{Mariam} \snm{Mussbah}}
\author[
addressref={aff2},
email={kkirev@nt.tuwien.ac.at}
]{\inits{KK}\fnm{Kiril} \snm{Kirev}}
\author[
   addressref={aff1,aff2},
   email={rnissel@nt.tuwien.ac.at}
]{\inits{RN}\fnm{Ronald} \snm{Nissel}}
\author[
	addressref={aff1,aff2},
	email={sschwarz@nt.tuwien.ac.at}
]{\inits{SS}\fnm{Stefan} \snm{Schwarz}}
\author[
	addressref={aff2},
	email={mrupp@nt.tuwien.ac.at}
]{\inits{MR}\fnm{Markus} \snm{Rupp}}


\address[id=aff1]{
  \orgname{Christian Doppler Laboratory for Dependable Wireless Connectivity for the Society in Motion} 
}
\address[id=aff2]{%
  \orgname{Institute of Telecommunications, TU Wien},
  \street{Gu\ss hausstra\ss e 25/389},
  \postcode{1040}
  \city{Vienna},
  \cny{Austria}
}


\begin{artnotes}
\end{artnotes}



\begin{abstractbox}

\begin{abstract} 
Research and development of mobile communications systems require a detailed analysis and evaluation of novel technologies to further enhance spectral efficiency, connectivity and reliability.
Due to the exponentially increasing demand of mobile broadband data rates and challenging requirements for latency and reliability, mobile communications specifications become increasingly complex to support ever more sophisticated techniques.
For this reason, analytic analysis as well as measurement based investigations of link level methods soon encounter feasibility limitations.
Therefore, computer aided numeric simulation is an important tool for investigation of wireless communications standards and is indispensable for analysis and developing future technologies.
In this contribution, we introduce the Vienna \acs{5G} Link Level Simulator, a \matlab{}-based link level simulation tool to facilitate research and development of \acs{5G} and beyond mobile communications.
Our simulator enables standard compliant setups according to \acs{4G} \acl{LTE}, \acs{5G} \acl{NR} and even beyond, making it a very flexible simulation tool.
Offered under an academic use license to fellow researchers it considerably enhances reproducibility in wireless communications research.
We give a brief overview of our simulation platform and introduce unique features of our link level simulator in more detail to outline its versatile functionality.
\end{abstract}




\end{abstractbox}
\end{fmbox}

\end{frontmatter}



\acresetall
\acuse{4G,5G,VLLS,UE} 

\section{Introduction} \label{sec:introduction}
Link level measurements, analysis and simulation are fundamental tools in the development of novel wireless communication systems, each offering its own unique benefits.
Measurements provide the basis for channel modeling and are thus a prerequisite of analysis and simulations.
Furthermore, measurement-based experiments present the ultimate performance benchmark of any transceiver architecture and are therefore indispensable in the system development cycle.
However, performing measurements is very costly, time consuming and hard to adapt to specific communication scenarios; hence, their application is commonly kept at a necessary minimum.
Compared to that, link level simulations facilitate rapid prototyping and comparison of competing technologies, enabling to gauge the potential of such technologies early on in the research and development process.
A big advantage of analytic investigations is their potential to reveal pivotal relationships amongst key parameters of a system; yet, analytic tractability often requires application of restrictive assumptions and simplifications, limiting the value of analytic results under realistic conditions.
To investigate highly complex systems, such as wireless transceivers, and to efficiently evaluate the performance of novel technologies, link level simulations are thus often the preferred method of choice, enabling the incorporation of realistic and practical constraints/restrictions, which in many cases significantly alter the picture drawn by purely analytic investigations.
Nevertheless, only by complementing measurements, analysis and simulations it is possible to reap the benefits of all three approaches.  

In this paper, we present the newest member of our suite of \ac{VCCS}: the \ac{VLLS}.
Our mobile communications research group at the Institute of Telecommunications at TU Wien has a long and successful history of developing and sharing standard-compliant cellular communications simulators under an academic use license, with the goal of enhancing reproducibility in wireless communications academic research~\cite{Meh11a,vccsurl}.
The implementation of our work-horse of the past nine years, that is, the Vienna LTE Simulators~\cite{VLS-2016}, started back in 2009, leading to three reliable standard compliant LTE simulators: a downlink system level simulator~\cite{Iku10,Tar15} and two link level simulators, one for uplink~\cite{Zoe16} and one for downlink~\cite{Sch13}.
Today the Vienna LTE Simulators count more than 50\,000 downloads in total.
Even though the path of \ac{3GPP} towards \ac{5G} is largely based on \ac{LTE}, we soon recognized that evolving our \ac{LTE} simulators towards \ac{5G} is not straightforward due to a lack of flexibility of the simulation platform in terms of implementation and functionality.
Mobile communications within \ac{5G} is expected to support much more heterogeneous and versatile use cases, such as \ac{EMBB}, \ac{MMTC} or \ac{URLLC}, as compared to \ac{4G}.
Furthermore, a multitude of novel concepts and contender technologies within \ac{5G}, e.g., full-dimension/massive MIMO beamforming~\cite{Lu2014,Larsson2014,Ji2016}, mixed numerology multicarrier transmission~\cite{Zaidi2016,Guan2017}, non-orthogonal multiple access~\cite{Ali2017,Ding2017}, and transmission in the \ac{MMW} band~\cite{Heath2016,Roh2014}, requires careful identification, abstraction and modeling of key parameters that impact the performance of the system.
We therefore decided to rather invest the required implementation effort into new simulators to extend our \ac{VCCS} simulator suite and to evolve to the next generation of mobile communications with dedicated \ac{5G} link and system level simulators.

The \ac{VLLS} focuses on the \ac{PHY} of the communication system. Correspondingly, the scope is on point-to-point simulations of the transmitter-receiver chain (channel coding, \ac{MIMO} processing, multicarrier modulation, channel estimation, equalization,...) supporting a broad range of simulation parameters.
Nevertheless, multi-point communications with a small number of transmitters and receivers are possible (limited only by computational complexity) to simulate, e.g., multi-point precoding techniques~\cite{Sch14b}, rate splitting approaches~\cite{Clerckx2016}, interference alignment concepts~\cite{Zhao2016}.
The transmission of signals over wireless channels thereby is implemented up to the individual signal samples and thus provides a very high level of detail and accuracy.

The \ac{VLLS} models the shared data channels of both, uplink and downlink transmissions, that is, the \ac{PDSCH} and the \ac{PUSCH}.
The simulator is implemented in \matlab{} utilizing object oriented programming methods and its source code is available for download under an academic use license~\cite{vccsurl}; hence, we attempt to continue our successful approach of facilitating reproducible research with this unifying simulation platform.
The simulator allows for (and includes) parameter settings that lead to standard compliant systems, including \ac{LTE} as well as \ac{5G}.
Yet, the versatile functionality of the simulator provides the opportunity to go far beyond standard compliant simulations, enabling, e.g., the evaluation of a cornucopia of different combinations of \ac{PHY} settings as well as the co-existence investigation of candidate \ac{5G} technologies.
The \ac{VLLS} acts in close orchestration with its sibling the Vienna \ac{5G} \ac{SL} Simulator: the \ac{LL} simulator is employed to determine the \ac{PHY} abstraction models utilized on \ac{SL} to facilitate computationally efficient simulation of large-scale mobile networks.

\section{Scientific Contribution} \label{sec:comparison}
Computer aided numerical simulations are a well established tool for analysis and evaluation of wireless communications systems.
Hence, a number of commercial and academic link level simulators is offered online.
Amongst these tools, our new \ac{VLLS} supports a variety of unique features as well as a highly flexible implementation structure that allows for easy integration of additional  components.
In this section, we provide an overview of existing similar simulation platforms and compare them to our new simulator.
We thereby restrict to academic tools that are, as our simulator, available for free, and leave aside commercial simulators.
Further, we point out the distinct features offered by the \ac{VLLS} and thereby state our scientific contribution.

\subsection{Related Work - Existing Simulation Tools} \label{sec:related_work}
With each new wireless communications standard, the need for simulation emerges.
This is necessary for the evaluation, comparison and further research as well as the development of communication schemes, specified within a certain standard.
The existence of various simulation tools for the \ac{PHY} of \ac{LTE}, such as~\cite{piro2011simulating,bultmann2009openwns} or the Vienna \acs{LTEA} \ac{LL} Simulator~\cite{Sch13,Zoe16,VLS-2016}, developed by our research group, supports this claim. 
These tools, however, are mainly based on \ac{3GPP} Release 8 to Release 10 and do not offer features and functionality specified and expected for \ac{5G}.
There exist some commercial products, that is, link level simulators, that claim to support simulation of \ac{5G} scenarios.
However, as we aim to support academic research also beyond the currently standardized features, we compare our \ac{VLLS} to other freely available academic simulators only.
An overview of existing \ac{LL} simulation tools is provided in~\cref{tab:simulators}.

\begin{table*}[h!]
	\caption{Overview of existing \acs*{LL} simulators.}
	\label{tab:simulators}
	\begin{tabular}{l|cc p{2.8cm} p{2.5cm} P{2.5cm}}
		\hline \hline
		\diagbox[width=35mm]{\textbf{simulator}}{\textbf{property}}	& \textbf{language/platform}	& \textbf{multi-link} 	& \textbf{waveforms}					& \textbf{channel codes} 									& \textbf{flexible numerology} \\
		\hline
		GTEC 5G LL Simulator 										& \matlab{}						& no					& \acs*{OFDM},\acs*{FBMC}				& -																						& no \\
		ns-3	 													& C++, Python					& yes					& - (\acs*{PHY} model)					& - (\acs*{PHY} model)	 																& no \\
		OpenAirInterface 											& C 							& yes 					& \acs*{OFDM} 							& turbo 																				& no \\
		\ac{VLLS}													& \matlab{}						& yes					& \acs*{CPOFDM},\acs*{FOFDM},\newline \acs*{FBMC},\acs*{UFMC},\acs*{WOLA} 	& \acs*{LDPC}, turbo,\newline polar, convolutional 	& yes \\
		\hline \hline
	\end{tabular}
\end{table*}

The \emph{GTEC 5G \ac{LL} Simulator} is an open source link level simulator developed at the University of A Coru\~na~\cite{dominguez2016gtec}.
It is based on \matlab{} and offers highly flexible implementation based on modules.
By implementation of new modules, it is even possible to simulate different wireless communications standards, such as \acs{WIMAX} or \ac{5G}.
The current \ac{5G} module offers two \ac{PHY} transmission waveforms, namely \ac{OFDM} and \ac{FBMC}.
\Ac{FEC} channel coding is not supported.
Further, this simulator focuses on single-link performance and does neither support multi-user nor multi-transmitter (multi \ac{BS}) scenarios.
Simulating \ac{IUI} of non-orthogonal users, e.g., \ac{NOMA} or mixed numerology use-cases, is not possible with this tool.

A well known tool for simulation of communications networks is the \emph{ns-3 simulator}~\cite{henderson2008network}, which is the successor of the \emph{ns-2 simulator}.
Although this tool has to be understood as a set of open source modules forming a generic network simulator, there exists an \ac{LTE} module~\cite{piro2011lte} that allows to simulate \ac{4G} networks.
Further, there exists a module~\cite{mezzavilla20155g,mezzavilla2017end} that covers \ac{MMW} propagation aspects of \ac{5G}.
Still, the main focus of the \emph{ns-3 simulator} lies on network simulations.
The mentioned \ac{LTE} module models radio resources with a granularity of resource blocks and does not consider time signals on a sample basis, as our simulator does.
Therefore, the \emph{ns-3 simulator} does not provide the detailed level of \ac{PHY} accuracy that distinguishes pure link level simulation tools from system and network level simulators, where a certain degree of physical layer abstraction is unavoidable to manage computational complexity.

The \emph{OpenAirInterface} is an open source platform offered by the Mobile Communications Department of EURECOM~\cite{nikaein2014openairinterface}.
Currently the implementation is based on \ac{3GPP} Rel. 8 and supports only parts of later releases.
The platform offers flexibility that allows for simulation of aspects of future mobile communications standards, such as \ac{CRAN} or \ac{SDN}.
The simulation platform supports the simulation of the core network as well as the \ac{RAN} and considers the complete protocol stack from the \ac{PHY} to the network layer.
It offers two modes for \ac{PHY} emulation, where the more detailed mode even considers actual transmission of signals over emulated channels.
Still, this simulator focuses on simulation of networks in terms of a complete protocol stack implementation.
However, details of \ac{PHY} transmissions, such as waveform, channel coding, numerology or reference signals for channel estimation and synchronization, are not considered.

Accurate \ac{LL} simulations require sophisticated channel models that realistically represent practically relevant propagation environments.
Since \ac{5G} introduces novel \ac{PHY} technologies, such as, full-dimension \ac{MIMO} and transmission in the \ac{MMW} band, also channel models need to be updated and revised.
The modular implementation structure of our simulator supports easy integration of dedicated wireless channel models and emulators, such as~\cite{jaeckel2014quadriga} or~\cite{samimi20163,sun2017novel}.

\subsection{Scientific Contribution and Novelty of our Simulator}
The currently ongoing evolution from \ac{LTE} towards \ac{5G} shows that \ac{LL} simulations are still a very active research topic, since many different candidate \ac{RAN} and \ac{PHY} schemes need to be gauged and compared against each other.
The \ac{VLLS} supports these needs and allows for a future-proof evaluation of \ac{PHY} technologies due to its versatility.
The simulator allows almost all \ac{PHY} parameters to be chosen freely, such that any multi-carrier system can be simulated; specifically, by setting parameters according to standard specifications, it is possible to conduct standard-compliant simulation of \ac{LTE} or \ac{5G} (we provide corresponding parameter files in the simulator package).
Due to the modular structure and application of object oriented programming, further functionality, such as additional channel models, can easily be included.

Our \ac{LL} simulator focuses on simulating \ac{PHY} effects in a high level of detail.
It considers the actual transmission of time signals over emulated wireless channels in a granularity of individual samples.
This allows the detailed analysis of \ac{PHY} schemes of current and future mobile communications systems, e.g., investigating the impact of the channel delay and Doppler spread on various \ac{PHY} waveforms and numerologies (see~\cref{sec:flexible_numerology}).

The following specific aspects distinguish our \ac{VLLS} from the tools summarized in~\cref{sec:related_work}.
\begin{itemize}
	\item \emph{\ac{PHY} methods even beyond \ac{5G}:}
	As mentioned above, the simulator supports standard-compliant simulation of the \ac{PDSCH}/\ac{PUSCH} of \ac{LTE} and \ac{5G} by implementing the signal processing chains described in~\cite{3GPP_TS_36211,3GPP_TS_38211}.
	Yet, simulation parameters of \ac{AMC}, \ac{MIMO} processing and baseband multicarrier waveforms are not restricted to standard-compliant values.
	In addition, the modular simulator structure allows for easy integration of novel functionality, such as additional waveforms or \acp{MCS}, to investigate candidate technologies of future mobile communication systems.
	As an example, we have implemented FBMC transmission to support comparison with the filtered/windowed \ac{OFDM}-based waveforms considered within 5G standardizations (\ac{WOLA}, \ac{UFMC}, \ac{FOFDM}). 
	\item \emph{Flexible numerology:}
	As introduced by \ac{3GPP} for \ac{5G}, the concept of flexible numerology describes the possibility to adapt the time and frequency span of a resource element.
	This means, that the subcarrier spacing and the symbol duration of the multicarrier waveform are adaptable to support a variety of service requirements (latency, coverage, throughout), channel conditions (delay or Doppler spread) and carrier frequencies.
	As these parameters are freely adjustable in our simulator, effects of different numerologies are investigatable, even beyond the standardized range~\cite{3GPP_TS_38211}.
	To summarize, the simulator enables comparison and optimization of numerologies of several multicarrier waveforms (see~\cref{sec:modulation}) in combination with various channel codes (see~\cref{sec:coding}) under arbitrary channel conditions in terms of delay and Doppler spread.
	\item \emph{Multi-link simulations:}
	The \ac{VLLS} is capable of simulating multiple \acp{UE} and \acp{BS} (only restricted by computational complexity).
	Although analysis of large networks with a high number of \acp{UE} and \acp{BS} is not the goal of a link level simulation, this feature allows investigation of \ac{IUI}.
	While this feature was not required for \ac{LTE}'s \ac{LL}, since user signals within a cell were automatically orthogonal due to the application of \ac{OFDM} with a fixed numerology, it is interesting in the context of \ac{5G} as users with different numerologies are not orthogonal anymore.
	The \ac{VLLS} enables the investigation of \ac{IUI} in such mixed numerology use-cases.
\end{itemize}

\section{Simulator Structure} \label{sec:structure}
In this section we provide a short general description of the \ac{VLLS} and a brief introduction of the simulator's structure.
In addition to this overview, supplementary documents, such as a user manual as well as a detailed list of features, are provided on our dedicated simulator web page~\cite{vccsurl}.

Link level simulations in most cases assume a fixed \ac{SNR} for the transmission link between transmitter and receiver.
We slightly deviate from this common approach in our simulator, since we support multiple different waveforms that achieve different \acp{SNR} for a given total transmit power.
Hence, rather than fixing the \ac{SNR}, we fix the transmit and noise power and determine the \ac{SNR} as a function of the applied waveform.
Additionally, since we support multi-link  transmissions, we introduce individual path loss parameters for these links, to enable controlling the SIR of the individual connections.
However, in contrast to a SL simulator, we do not introduce a spatial network geometry to determine the path loss, but rather set the path loss as an input parameter of the simulator.
The goal of \ac{LL} simulations is to obtain results in terms of \ac{PHY} performance metrics, such as throughput, \ac{BER} or \ac{FER}, which are representative for the average system performance within the specified scenario.
To this end,  Monte Carlo simulations are carried out and results are averaged over a certain number of channel, noise and data realizations.
To gauge the statistical significance of the obtained results, the simulator calculates the corresponding 95 percentile confidence intervals.

\begin{figure}
	\centering
	\includegraphics[width=76mm]{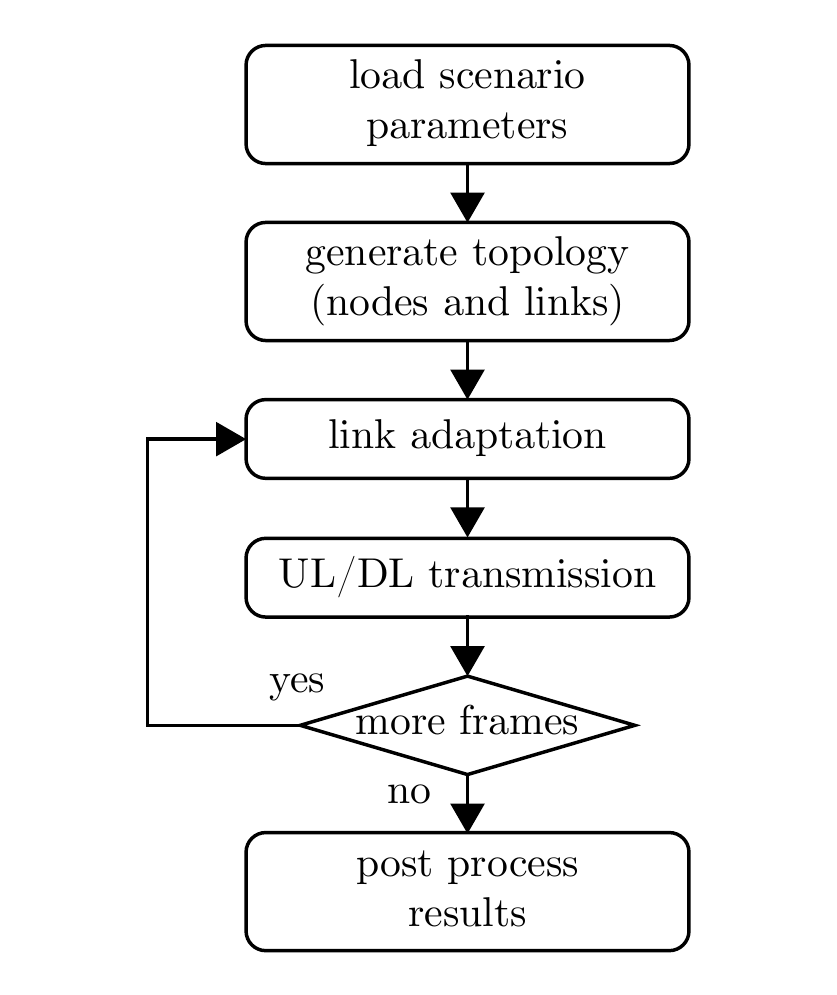}
	\caption{Basic structure of the simulator.}
	\label{fig:structure_flowchart}
\end{figure}

\cref{fig:structure_flowchart} illustrates the basic processing and simulation steps applied by the simulator. 
The initial step is to specify and supply a scenario file to the simulator.
The file contains all the information necessary for the simulation.
Setting up a scenario begins with the specification of the network topology, that is, defining all the nodes and their associated connections in the network.
The nodes take the role of either a \ac{BS} or a \ac{UE}. Arbitrarily meshed connections between these nodes are supported. The connections can serve as downlink, uplink, or sidelink (device-to-device link). Furthermore, inter- and intra-cell interference is easily captured, as it only requires to establish the corresponding connections between the supposedly interfering nodes.

Every connection is represented in the simulator by a so-called link object.
It is the most fundamental building block of the simulator and contains all the \ac{PHY} functionality objects, such as, channel coding, modulation, \ac{MIMO} processing, channel generation and estimation, \ac{CSI} feedback calculation, etc. Moreover, the link object also contains the generated signals throughout the whole transceiver chain of the specific connection.

After the network topology is specified, the next step in the scenario setup is to enter the transmission parameters.
This covers the whole transmission chain, including the specification of the applied channel coding scheme, the multicarrier waveform, the applied channel model, as well as the equalizers and decoders employed by the receiver.
Parameters can either be set locally for each link and node, or conveniently globally, in case all links and nodes use the same settings.
Setting different parameters for different links enables coexistence investigations of multiple technologies; for example, one cell could be set up to operate with \ac{OFDM} and turbo coding while the other cell uses \ac{FBMC} with \ac{LDPC} coding.
This allows to investigate the sensitivity w.r.t. out-of-cell interference of such systems.
Notice, since signals are processed on a sample basis, the modeling of such interference is highly accurate.

Once the scenario file is ready, it gets loaded by the main script of the simulator, where the simulation is set up according to the input topology and parameters.
The simulation is carried out on a frame-by-frame basis over a specified sweep parameter, such as the path loss, transmit power, or velocity\footnote{Please note, that the velocity determines the maximum Doppler shift of the \ac{UE}'s channel. As there is no geometry in an \ac{LL} simulator, the user has no physical position that changes over time.}.
Within the simulation, the simulator performs full down- and uplink operation of all specified transmission links, including the possibility to activate \ac{LTE} compliant \ac{CSI} feedback as well as link adaptation in terms of \ac{AMC} and standard-compliant \ac{MIMO} processing (see~\cref{sec:feedback}).
The results for all nodes in the form of throughput, \ac{FER}, and \ac{BER} versus the sweep parameter (e.g., \acs{SNR}) are provided as simulator output.
In addition to these aggregated results, the simulator also stores simulation results of individual frames, to support further post-processing by the researcher.
The overall procedure is optimized in such a way that the overhead of the exchanged information during the simulation is minimal, and the operations are executed efficiently.
Moreover, parallelism can be enabled over the loop of the sweep parameter, which offers a substantial reduction in simulation time when run on multi-processor machines.

\section{Features} \label{sec:features}
In this section, we provide a more detailed description of the \ac{VLLS}.
The main components and features are described, giving insights in the available versatile functionality.
To highlight features, that make our \ac{LL} simulator unique, we further provide and discuss results of exemplary simulations.
All of these example scenarios are included with the simulator download package and are straightforward to reproduce.

\subsection{Channel Coding} \label{sec:coding}
The first processing block in the transmission chain is channel coding, where error correction and detection capability is provided to the transmitted signal.
The simulator supports the four coding schemes of convolutional, turbo, \ac{LDPC}, and polar codes.
These schemes were selected by \ac{3GPP} as the candidates for \ac{5G}, due to their excellent performance and low complexity state-of-the-art implementation.
\cref{tab:codingTable} summarizes the supported channel coding schemes and their corresponding decoding algorithms.

{\renewcommand{\arraystretch}{1.5}
	\begin{table}[ht!]
		\caption{Supported channel coding schemes.}
		\label{tab:codingTable}
		\begin{tabular}{l| m{2.3cm} |m{2.5cm}}
			\hline \hline
			\textbf{scheme}	& \textbf{construction/}\newline \textbf{encoding} & \textbf{decoding}\newline \textbf{algorithms} \\
			\hline
			turbo 			& \acs*{LTE}										& Log-MAP\newline Linear-Log-MAP\newline MAX-Log-MAP\\
			\acs*{LDPC}		& \acs*{5G} \acs*{NR} 								& Sum-Product\newline PWL-Min-Sum\newline Min-Sum\\
			polar 			& currently custom 									& SC\newline List-SC\newline CRC-List-SC \\
			convolutional 	& \acs*{LTE} 										& Log-MAP\newline MAX-Log-MAP \\
			\hline \hline
		\end{tabular}
	\end{table}
}

The turbo and convolutional codes are based on the \ac{LTE}~\cite{3GPP_TS_36212} standard, the \ac{LDPC} code follows the \ac{5G} \ac{NR}~\cite{3GPP_TS_38212} specifications, and for polar codes we currently use the custom construction in~\cite{Tah17} concatenated with an outer \ac{CRC} code.
This includes both the construction of the codes and also the whole segmentation and rate matching process as defined in the standards. 

The decoding of convolutional and turbo codes is based on the \(\log\)-domain \acs{BCJR} algorithm~\cite{Bahl1974}, that is, the Log-MAP algorithm, and its low complexity variants of MAX-Log-MAP~\cite{Koch1990} and Linear-Log-MAP~\cite{Cheng2000}.
For the \ac{LDPC} code, the decoder employs the Sum-Product algorithm~\cite{MacKay1999}, and its approximations of the Min-Sum~\cite{Chen2005} as well as the double piecewise linear PWL-Min-Sum~\cite{Mansour2003}.
The LDPC decoder utilizes a layered architecture where the column message passing schedule in \cite{Radosavljevic2005} is applied.
This allows for faster convergence in terms of the decoding iterations.
As for polar codes, the decoder is based on the \(\log\)-domain \ac{SC}~\cite{Arikan2009}, and its extensions of List-\ac{SC} and \ac{CRC}-aided List-\ac{SC}~\cite{Tal2011}.

In the remainder of this section, we consider an example simulation on channel coding performance with short block length.
The study of channel codes for short block lengths in combination with low code rates, is of interest in many applications.
Typical examples are the control channels of cellular systems, and the \ac{5G} \ac{MMTC} and \ac{URLLC} scenarios.
Here, we investigate the aspects of such combination using convolutional, turbo, \ac{LDPC}, and polar codes.
In order to do that, we need to have a complete freedom in setting the parameters of the block length and code rate.
Thanks to the modular structure of the simulator, we can use the channel coding object in a standalone fashion, thus eliminating the restrictions imposed by the other parts of the transmission chain, such as the number of scheduled resources or the target code rate for given channel conditions. 
\cref{tab:shotBlocksExampleTable} lists the simulation parameters of our setup.
For the decoding iterations and list size, we employ relatively large values to maximize the performance of the decoding algorithm.

\begin{table*}[h]
	\caption{The simulation parameters of the channel coding for short block lengths example. Three different channel codes are compared for the same short block length.}
	\label{tab:shotBlocksExampleTable}
	\begin{tabular}{l|c|c|c|c}
		\hline \hline
		\textbf{parameter} 		& \multicolumn{4}{c}{\textbf{value}} \\ \hline
		channel code		    & convolutional     & turbo   & \acs*{LDPC} & polar\\
		decoder 		        & MAX-Log-MAP 		& Linear-Log-MAP & PWL-Min-Sum & \acs*{CRC}-List-\acs*{SC} 	\\
		iterations/list size    & -					& 16 & 32 & 32    \\ \hline
		block length 			& \multicolumn{4}{c}{64 bits (48 info + 16 \acs*{CRC})} \\
		code rate				& \multicolumn{4}{c}{1/6} \\ 
		modulation				& \multicolumn{4}{c}{4\,\acs*{QAM}} \\ 
		channel 				& \multicolumn{4}{c}{\acs*{AWGN}} \\
		\hline \hline
	\end{tabular}
\end{table*}

For the \ac{LDPC} code, filler bits were added to the input block.
This is necessary to compensate the mismatch between the chosen block length and the dimensions of the \ac{5G} \ac{NR} parity check matrix.
However, the addition of filler bits reduces the effective code rate, since it results in a longer output codeword.
For this reason, we further puncture the output in such a way that the target length and code rate are met.
This might have a negative impact on the performance of the \ac{LDPC} code, as some parts of the codeword belong to the extra filler bits which are discarded after decoding.
Nonetheless, by following this procedure we guarantee that all the schemes are running with the same code rate.

\begin{figure}
	\centering
	\includegraphics[width=76mm]{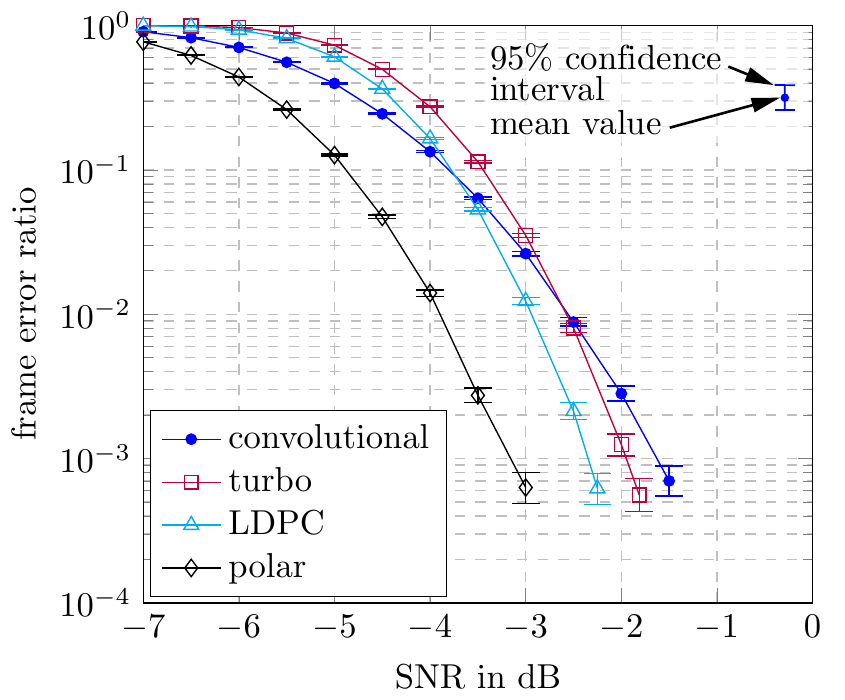}
	\caption{Simulation results for the channel coding example simulation. The \ac{FER} performance for small block lengths is given. We compare convolutional, turbo, \acs*{LDPC}, and polar codes for a block length of \(64\) bits and a code rate of \(1/6\), operating over the \acs*{AWGN} channel with 4\,\acs{QAM} signaling.}
	\label{fig:shortlblocks_fer}
\end{figure}

\cref{fig:shortlblocks_fer} shows the \ac{FER} performance of the aforementioned coding schemes.
It can be seen that the polar code has the best performance in such setup.
Namely, at the \ac{FER} of \(10^{-2}\), the polar code is leading by \(1\)\,dB against the \ac{LDPC} code, and by \(1.5\)\,dB against the turbo and convolutional codes.
At the lower regime of the \ac{FER}, the gap appears to get narrower.
Still, the polar code remains the clear winner.
This makes it an attractive choice for the scenarios of short block length.
However, other factors which are not considered here, such as decoding latency or hardware implementation aspects, influence the choice of a coding scheme.

\subsection{Modulation} \label{sec:modulation}
\begin{figure}
	\centering
	\includegraphics[width=76mm]{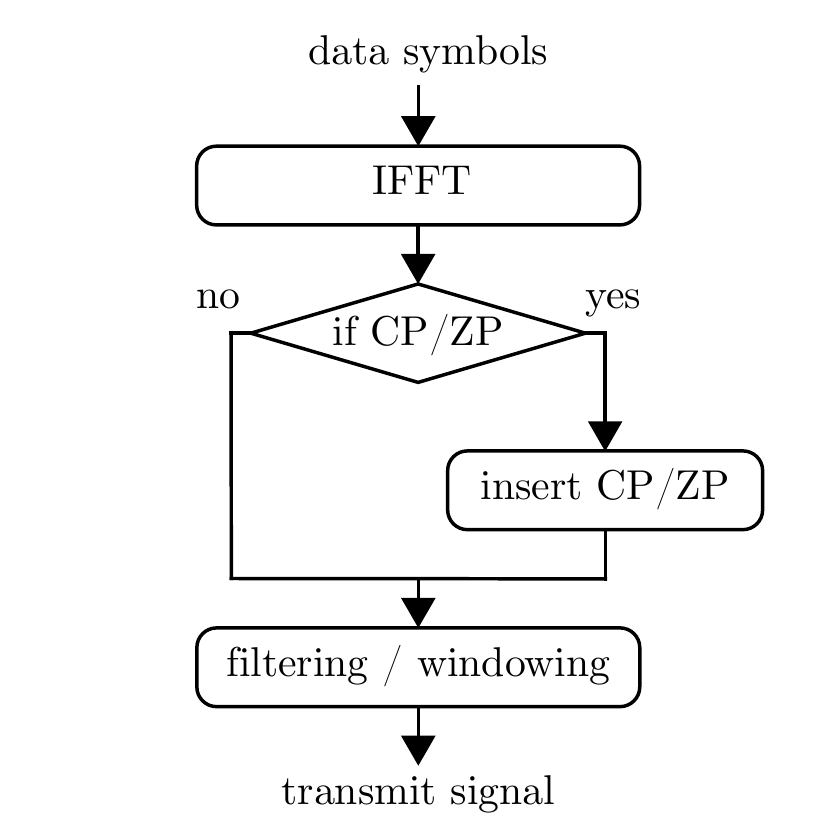}
	\caption{Generation of the transmit signal. This general principle is applicable for all supported multicarrier schemes.}
	\label{fig:transmitter_flowchart}
\end{figure}

\begin{figure}
	\centering
	\includegraphics[width=76mm]{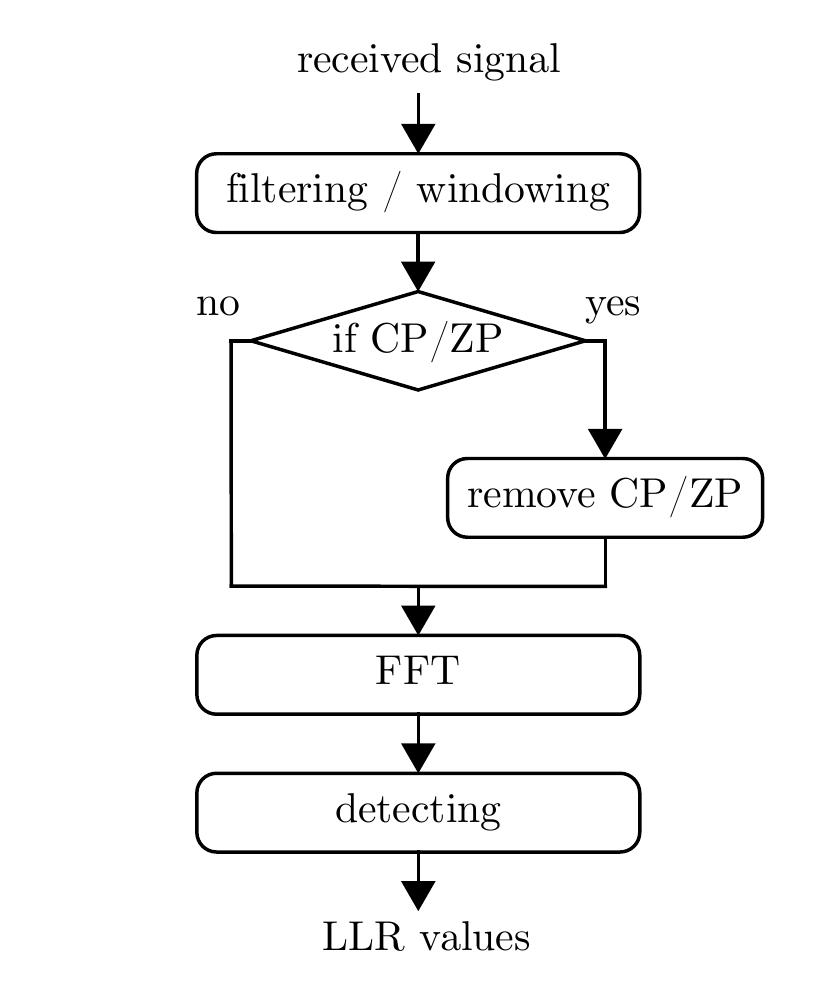}
	\caption{Processing of the received signal. This general principle is applicable for all supported multicarrier schemes.}
	\label{fig:receiver_flowchart}
\end{figure}

According to the current \ac{3GPP} specifications related to the \ac{NR} \ac{PHY} design, equipment manufacturers are unconstrained in the choice of \ac{OFDM}-based multicarrier waveforms~\cite{3GPP_TR_38912}.
\Ac{CPOFDM} will be the baseline multicarrier transmission scheme applied in \ac{5G}.
However, to reduce \ac{OOB} emissions and improve spectral confinement, manufacturers are free to add windowing or filtering on top of \ac{CPOFDM}.
Our simulator offers the versatility to support various multicarrier waveforms.
Besides \ac{OFDM}-based waveforms, such as, \ac{CPOFDM}, \ac{WOLA}, \ac{UFMC}, \ac{FOFDM}, we additionally support \ac{FBMC} as a promising candidate for the next generations beyond \ac{5G}~\cite{Nis17d}.
In the following we provide a brief description of each waveform supported by our simulator.
The basic signal flowcharts applied for signal generation and reception at the transmitter and receiver, respectively, for all of these waveforms are shown in~\cref{fig:transmitter_flowchart,fig:receiver_flowchart}.
In general, the filtering and windowing operations are applied in time domain, after the \ac{IFFT}.
Depending on the modulation scheme, these operations are performed per subband (\ac{UFMC}), per subcarrier (\ac{WOLA},\ac{FBMC}) or on the whole band (\ac{FOFDM}).
Some of the schemes employ \ac{CP}, such as \ac{CPOFDM}, \ac{WOLA} and \ac{FOFDM}, or \ac{ZP}, such as \ac{UFMC}, in order to prevent distortion caused by the multipath channel and by filtering or windowing.
To mitigate potential \ac{IUI} (or inter-subband interference), windowing/filtering is also applied at the receiver side~\cite{schaich2015subcarrier}.

\subsubsection{Cyclic Prefix OFDM}
Proposed in the downlink of the current \ac{LTE} system, \ac{CPOFDM} is currently the most prominent multicarrier waveform, since it is the standardized waveform not only of \ac{LTE} downlink, but also for WiFi 802.11~\cite{3GPP_TS_36201}.
It assumes a rectangular pulse of duration $T = T_{\rm{OFDM}} + T_{\rm{CP}}$ at the transmitter, where $T_{\rm{OFDM}}$ is the useful symbol duration and $T_{\rm{CP}}$ is the length of \ac{CP}.
This pulse shape enables very efficient implementation by means of an \ac{IFFT} at the transmitter side and an \ac{FFT} at the receiver side.
Unfortunately, the rectangular pulse also causes high \ac{OOB} emissions.
By applying a \ac{CP} the scheme avoids \ac{ISI} and preserves orthogonality between subcarriers in the case of highly frequency selective channels.
This simplifies equalization in frequency-selective channels, enabling the use of simple one-tap equalizers, but reduces the spectral efficiency at the same time due to the \ac{CP} overhead.

\subsubsection{Weighted Overlap and Add}
\Ac{WOLA} extends \ac{OFDM} by applying signal windowing in the time domain~\cite{zayani2016wola}.
Unlike conventional \ac{OFDM}, which employs a rectangular prototype pulse, \ac{WOLA} applies a window that smooths the edges of the rectangular pulse, improving spectrum utilization.
The window shape is based on the (root) raised cosine function determined by a roll-off factor that controls the windowing function.
Due to this windowing function, consecutive \ac{WOLA} symbols overlap in time.
This effect is compensated for by extending the length of the \ac{CP}.
In that way the scheme preserves the orthogonality of symbols and subcarriers, but it increases the overhead of the \ac{CP} and therefore reduces the spectral efficiency compared to \ac{OFDM}.
At the receiver side, windowing in combination with an overlap and add operation further reduces \ac{IUI}~\cite{Nis17d}.

\subsubsection{Universal Filtered Multicarrier}
As an alternative to windowing, filtering can be applied to \ac{OFDM} waveforms to reduce out-of-band emissions.
\Ac{UFMC} employs subband-wise filtering of \ac{OFDM} and is therefore an applicable waveform for \ac{5G} \ac{NR}~\cite{schaich2014waveform}.
Compared to \ac{OFDM}, \ac{UFMC} provides better suppression of side lobes and supports more efficient utilization of fragmented spectrum.
In our simulator, we follow the transceiver structure proposed in~\cite{vakilian2013universal,schaich2014waveform}.
At the transmitter side we apply a subband-wise \ac{IFFT}, generating the transmit signal in time domain.
In order to reduce \ac{OOB} emissions we apply filtering on a set of contiguous subcarriers, so called subbands.
There are several criteria for the filer design.
In our simulator, we employ the Dolph-Chebyshev filter since it maximizes the side lobe attenuation~\cite{geng2015ufmc}; however, other filters can easily be implemented.
\Ac{UFMC} employs a \ac{ZP} in order to avoid \ac{ISI} on time dispersive channels, although there are only minor differences compared to the utilization of a \ac{CP}~\cite{venkatesan2016ofdm}.
To restore the cyclic convolution property similar to \ac{CPOFDM}, which enables low-complexity \ac{FFT}-based equalization, the tail of the receveid signal is added to its beginning  at the receiver side.

\subsubsection{Filtered OFDM}
Very similar to \ac{UFMC} is \ac{FOFDM}; however, unlike \ac{UFMC}, which applies filtering on a chunk of consecutive subcarriers, \ac{FOFDM} includes a much larger number of subcarriers which are generally associated to different use cases~\cite{abdoli2015filtered}.
\ac{FOFDM} employs filtering at both, transmitter and receiver side.
If the total \ac{CP} length is longer than the combined filter lengths, the scheme restores orthogonality in an \ac{AWGN} channel.
Hence, compared to \ac{CPOFDM}, the \ac{CP} length is generally longer and thus the overhead is increased.
However, in order to keep the overhead at a minimum, the method usually allows a small amount of self-interference.
The introduced interference is controlled by the filter length and the \ac{CP} length.
Currently our simulator supports a filter based on a sinc pulse which is multiplied by a Hanning window, yet any other filter can easily be incorporated.

\subsubsection{Filter-Bank Multicarrier}
Although \ac{3GPP} decided that \ac{FBMC} will not be employed in \ac{5G}, it still has many advantages compared to \ac{OFDM} and is thus a viable candidate for the next generations beyond \ac{5G}.
One of the most significant advantage is the low \ac{OOB} emission.
However, narrow subcarrier filters in the frequency domain imply overlapping of symbols in the time domain.
\ac{FBMC} does not achieve complex-valued orthogonality, but only orthogonality of real-valued signals.
Nevertheless, in combination with offset-\ac{QAM}, the same spectral efficiency as in \ac{OFDM} can be achieved.
Additionally, in the case of doubly-selective channels, the method is able to significantly suppress \ac{ISI} and \ac{ICI} using conventional equalizers, such as a \ac{MMSE} equalizer~\cite{Nis17b,Mar16}.
However, for channel estimation or in the case of \ac{MIMO} transmissions, imaginary interference has a more significant impact and requires a special treatment~\cite{Nis17c}.

\subsection{MIMO Transmissions}
The \ac{VLLS} supports arbitrary antenna configurations and various \ac{MIMO} transmission modes.
Not only may the number of transmit and receive antennas be set to any value, also this parameter is individually adjustable for each node.
The behavior of the \ac{MIMO} transmitter and receiver is selected via the transmission mode.
Currently the available options are transmit diversity, \ac{OLSM}, \ac{CLSM} and a custom transmission mode that is freely configurable.
The transmit diversity mode leads to a standardized version of Alamouti's space-time codes~\cite{3GPP_TS_36211}.
\ac{OLSM} and \ac{CLSM} both are \ac{LTE} standard compliant \ac{MIMO} transmission modes, where link adaptation is performed according to~\cref{sec:feedback}.
The additional custom transmission mode allows a flexible setting of parameters.
For this configuration, the number of active spatial streams, the precoding matrix as well as the \ac{MIMO} receiver are freely selectable.
Currently \ac{ZF}, \ac{MMSE}, sphere decoding and \ac{ML} \ac{MIMO} detectors are implemented.

\subsection{Feedback} \label{sec:feedback}

\begin{figure}
	\centering
	\includegraphics[width=76mm]{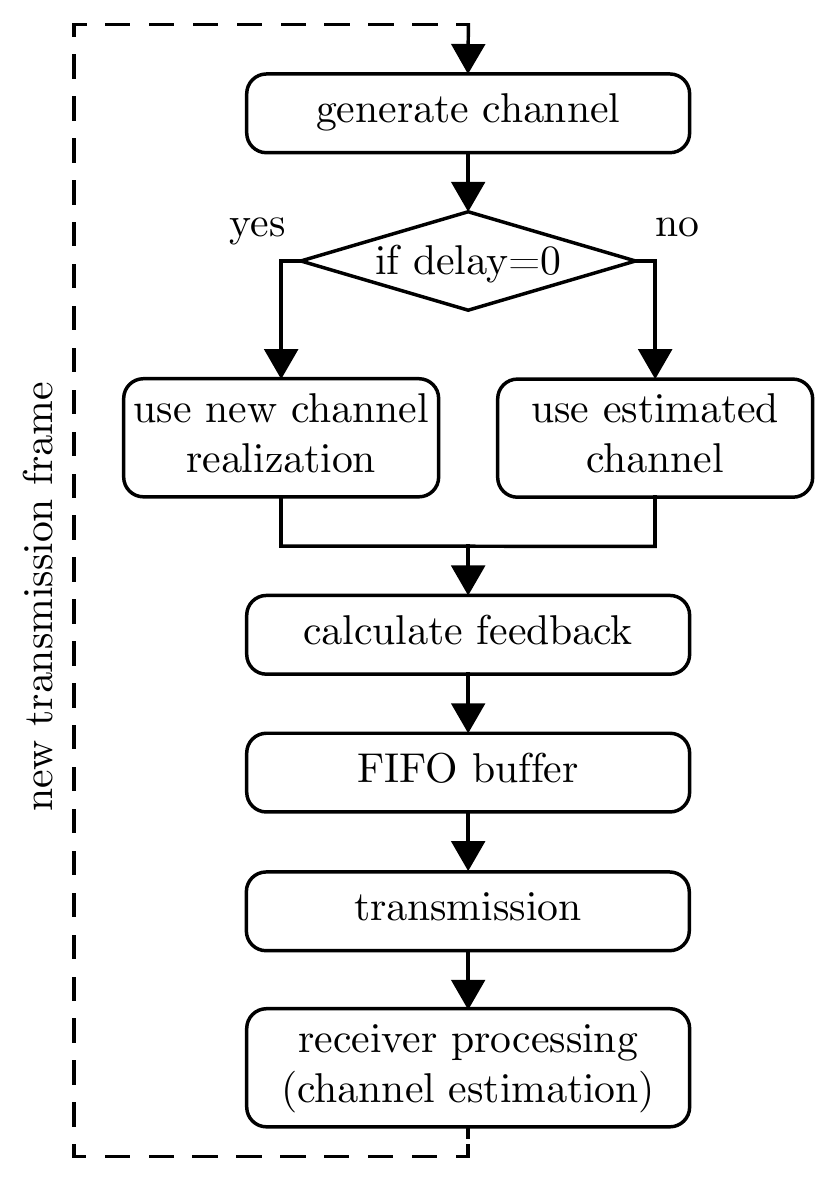}
	\caption{Feedback calculation flowchart. The delay of the feedback channel is implemented by means of a FIFO buffer.}
	\label{fig:feedback_flowchart}
\end{figure}

To adapt the transmission parameters to the current channel conditions, \ac{CSI} at the transmitter is required.
Since up- and downlink are implemented for \ac{FDD} mode, the  reciprocity of the channel cannot be exploited. For non-reciprocal channels, the receiver has to estimate the channel and then feed back the \ac{CSI} to the transmitter.
To reduce the overhead, the fed back \ac{CSI} is quantized.
The feedback calculation is an intelligent way of quantizing the \ac{CSI}, it comprises the \ac{PMI}, \ac{RI} and \ac{CQI}.
By the \ac{CQI} feedback the transmitter selects one of 15 \acp{MCS}.
By the \ac{PMI} the transmitter selects a precoding matrix from a codebook and the \ac{RI} informs the transmitter about the number of active transmission layers. 

The algorithm for calculation of the feedback parameters is based on~\cite{Sch10,Sch10a}.
To reduce the complexity, the feedback calculation is decomposed into two separate steps.
In the first step the optimum \ac{PMI} and \ac{RI} are jointly evaluated, maximizing the sum mutual information over all scheduled subcarriers.
In the second step we choose the \ac{CQI} with the largest rate that achieves a \ac{BLER} below a certain threshold. The \ac{CQI} value is found by mapping the post equalization \ac{SINR} of all scheduled subcarriers to an equivalent \ac{AWGN} channel \ac{SNR}.

\cref{fig:feedback_flowchart} shows a flowchart describing the feedback process.
The feedback channel is modeled as simple delay, which is implemented by means of a FIFO buffer of corresponding size; hence, we do not consider transmission errors in the feedback path, only the processing delay.
For a feedback delay of $n > 0 $, the estimated channel is used at the receiver for the feedback calculation.
The calculated feedback is then fed into the FIFO buffer.
For the first $n$ transmissions, all three feedback parameters are set to the default value 1.
The delay has to be sufficiently smaller than the coherence time to ensure similar channel conditions.
The feedback calculation is placed after the generation of the channel and before the transmission.
This enables simulations with instantaneous (zero delay) feedback as the newly generated channel is immediately available for the feedback calculation.  

For the \ac{CLSM} and \ac{OLSM} transmission modes, the feedback is configured automatically, whereas for the custom transmission mode the feedback is configured manually in the scenario file.
For all three transmission modes, the feedback delay and the type of averaging within the \ac{SINR} mapping have to be set in the scenario file.

\subsection{Channel Models}
As the aim of \ac{LL} simulation is acquisition of the average link performance, many random channel realizations are necessary per scenario.
There exists no network geometry and therefore no path-loss model.
A link's path-loss is an input parameter, determining the user's average \ac{SINR}.
Therefore, the channel model only includes small scale fading effects while its average power is dictated by the given path-loss.

The use of a universal spatial channel model, such as the QUADRIGA model~\cite{jaeckel2014quadriga} or the \ac{3GPP} 3D channel model~\cite{Ade16a,3GPP_TR_36873}, is of limited benefit due to the lack of geometry.
We offer frequency selective and time selective fading channel models.
The frequency selectivity is implemented as \ac{TDL} model.
Currently we offer implementations for to Pedestrian A, Pedestrian B, Vehicular A~\cite{3GPP_TR_25890}, TDL-A to TDL-C~\cite{3GPP_TR_38901}, Extended Pedestrian A and Extended Vehicular A~\cite{3GPP_TS_36104}.
To model the channel's time selectivity, the fading taps change over time to fit a certain Doppler spectrum.
A Jake's as well as an uniform Doppler spectrum are currently implemented.
Jake's model also supports time-correlated fading across frames according to a model from~\cite{zheng2003simulation} with a modification from the appendix of~\cite{zemen2005time}.
For time-invariant channels, the \ac{TWDP} fading model~\cite{durgin2002new} is employed, which is a generalization of the Rayleigh and Rician fading models.
In contrast to the Rayleigh fading model, where only diffuse components are considered, and the Rician fading model, where a single specular component is added, two specular components together with multiple diffuse components are considered in the \ac{TWDP} fading model.
The two key parameters for this model are $K$ and $\Delta$.
Similar to the Rician fading model, the parameter $K$ represents the power ratio between the specular and diffuse components.
The parameter $\Delta$ is related to the ratio between peak and average specular power and thus describes the power relationship between the two specular components.
By proper choice of $K$ and $\Delta$, the \ac{TWDP} fading model is able to characterize small scale fading for a wide range of propagation conditions, from no fading to hyper-Rayleigh fading.
\cref{tab:fading_parameters} shows typical parameter combinations and their corresponding fading statistic.
In contrast to classical models, the \ac{TWDP} fading model allows for destructive interference between two dominant specular components.
This enables for a possible worse than Rayleigh fading behavior, depending on the fading model parameters. 

Spatial correlation of \ac{MIMO} channels is implemented via a Kronecker correlation model with correlation matrices as described in~\cite{3GPP_TS_36101}.

\begin{table}[ht!]
	\caption{Parameters of the \acs*{TWDP} fading model. For the case of a time-invariant channel, the \acs*{TWDP} fading is parametrized by $K$ and $\Delta$.}
	\label{tab:fading_parameters}
	\begin{tabular}{c|c|c}
		\hline \hline
		\textbf{fading statistic} 	& $K$			& $\Delta$ 		\\ \hline
		no fading 					& $\infty$ 		& $0$ 			\\	
		Rician						& $>0$ 			& $0$ 			\\	
		Rayleigh 					& $0$ 			& - 			\\	
		Hyper-Rayleigh 				& $\infty$ 		& $\approx 1$ 	\\	
		\hline \hline
	\end{tabular}
\end{table}

\subsection{Flexible Numerology} \label{sec:flexible_numerology}
In this section, an example simulation scenario, demonstrating the concept of flexible numerology, is presented.
As previously emphasized, one of the key advantages of our simulator compared to other simulators is the support of flexible numerology.
According to \ac{3GPP}, there are three \ac{5G} use case families: \ac{URLLC}, \ac{MMTC} and \ac{EMBB}.
In order to meet different requirements of a specific use case, \ac{3GPP} proposes a flexible \ac{PHY} design in terms of numerology.
Flexible numerology refers to the parametrization of the multicarrier scheme.
It means that we are flexible to choose different subcarrier spacing and thus symbol and \ac{CP} duration according to the desired latency, coverage or carrier frequency.
On the other hand, the right choice of subcarrier spacing has to depend also on the channel conditions.
We investigate the impact of different channel conditions for different numerologies in~\cite{Mar18}.
Additionally, in~\cite{Mar18} we apply the optimal pilot pattern obtained accounting not only for the channel conditions, but also for the channel estimation error as a consequence of imperfect channel knowledge.
In this section we assume perfect channel knowledge and investigate the sensitivity of different numerologies w.r.t. the delay and Doppler spread of the channel.

\begin{table}[h!]
	\centering
	\caption{Parameters for the flexible numerology simulation example. Three different numerologies are employed.}
	\label{tab:table1}
	\begin{tabular}{l|ccc}
		\hline \hline
		\textbf{parameter} 				& \multicolumn{3}{c}{\textbf{value}} \\
		\hline
		subcarrier spacing 				& 15\,kHz 			& 60\,kHz	& 120\,kHz\\
		number of symbols per frame 	& 14				& 56		& 112 \\
		\ac{CP} duration 				& 4.76\,\textmu s 		& 1.18\,\textmu s	& 0.59\,\textmu s \\ \hline
		bandwidth                  		& \multicolumn{3}{c}{5.76\,MHz} \\
		carrier frequency        		& \multicolumn{3}{c}{5.9\,GHz} \\
		modulation alphabet       		& \multicolumn{3}{c}{64\,\acs{QAM}} \\
		channel model       			& \multicolumn{3}{c}{\acs*{TDL}-A}\\
		\hline \hline
	\end{tabular}
\end{table}

\begin{figure}
	\centering
	\includegraphics[width=76mm]{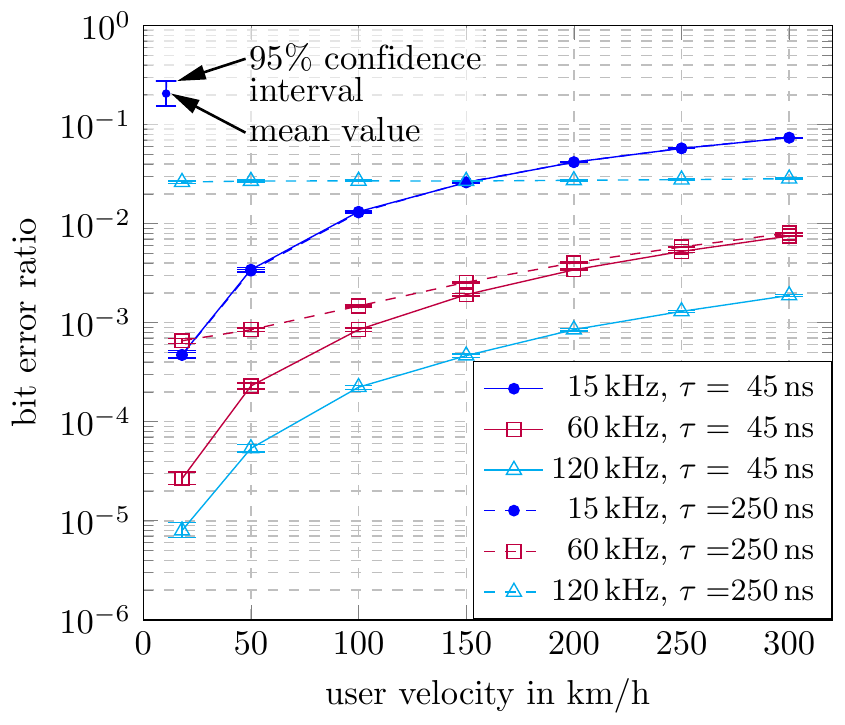}
	\caption{Simulation results of the flexible numerology example. The uncoded \acs*{BER} is considered for a sweep over user velocity for three different numerologies and two different channel \acs*{RMS} delay spreads.}
	\label{fig:BER}
\end{figure}

The parameters used for the simulations shown in~\cref{fig:BER} are given in~\cref{tab:table1}.
According to \ac{3GPP}, subcarrier spacings for \ac{5G} are scaled versions of the basic 15\,kHz subcarrier spacing by a factor $2^k$, where $k$ is an integer, in the range from $0$ up to $5$~\cite{3GPP_TS_38211}.
In this simulation example we take three values for subcarrier spacings - 15\,kHz, 60\,kHz and 120\,kHz.
We keep the same bandwidth of 5.76\,MHz using different subcarrier spacings, and vary the number of subcarriers and symbols proportionally.
We employ the \ac{TDL}-A channel model with two different \ac{RMS} delay spreads of $\tau=45\,$ns and $\tau=250$\,ns~\cite{3GPP_TR_38901}.
The channel's time selectivity is determined according to Jake's Doppler spectrum where the maximum Doppler frequency is given by the \ac{UE}'s velocity.
In~\cref{fig:BER} we show the behavior of the \ac{BER} versus \ac{UE} velocity, using the $5.9$\,GHz carrier frequency, which is typical for the vehicular communication scenarios.
We want to show the impact of the Doppler shift and different channel delay spread on the different subcarrier spacings.
In general, we observe that the \ac{BER} increases with \ac{UE} velocity due to growing \ac{ICI}.
In case of the short \ac{RMS} delay spread, the transmission achieves a lower \ac{BER} compared to larger subcarrier spacings, since the large subcarrier spacing is more robust to \ac{ICI} and also the \ac{CP} length is sufficient compared to the maximum channel delay spread.
On the other hand, with the long \ac{RMS} delay spread channels, \ac{ISI} is present with large subcarrier spacings due to insufficient \ac{CP} length.
In~\cref{fig:BER} \ac{ISI} is already pronounced with $60$\,kHz.
For $120$\,kHz, \ac{ISI} dominates over \ac{ICI} for the entire range of considered user velocities; hence, the \ac{BER} curve is flat.

\subsection{Multi-link Simulations}
To demonstrate the capability of simulations employing multiple communication links of our simulator, we show a further example simulation in this section.
Different users were inherently orthogonal due to the deployment of \ac{OFDM} in \ac{LTE}.
Due to the concept of mixed numerologies in \ac{5G} or the employment of non-orthogonal waveforms in future mobile communications systems, users are not necessarily orthogonal anymore.
As our simulator allows multiple communication links, it enables investigation of scenarios where users interfere with each other.

\begin{figure}
	\centering
	\includegraphics[width=76mm]{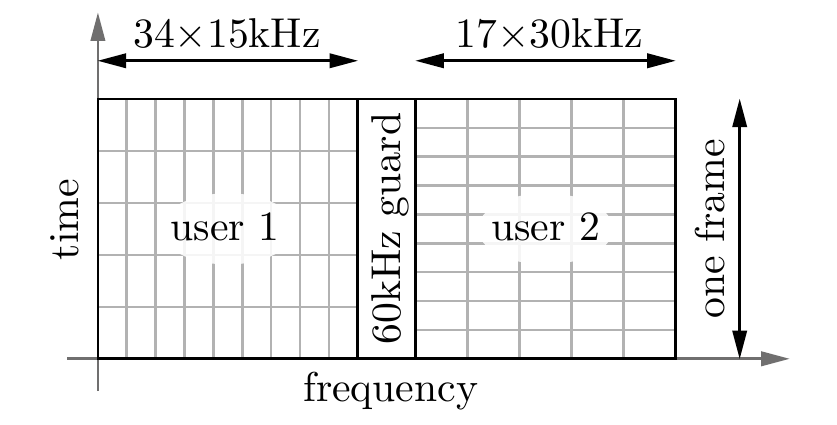}
	\caption{Schedule of the multi-link simulation scenario. The two scheduled users are not orthogonal as they employ different numerologies.}
	\label{fig:inter_user_interference_schedule}
\end{figure}

To demonstrate this feature, we consider an uplink transmission of two users with mixed numerology.
User\,1 employs a subcarrier spacing of $15\,$kHz while User\,2 employs $30\,$kHz which makes them non-orthogonal.
The users are scheduled next to each other in frequency as shown in~\cref{fig:inter_user_interference_schedule}.
The guard band of $60\,$kHz is intended to reduce the \ac{IUI} at the price of spectral efficiency.
We consider User\,1 as the primary user that suffers from interference generated by User\,2.
The \ac{BS}'s receiver is not interference aware.

\begin{figure}
	\centering
	\includegraphics[width=76mm]{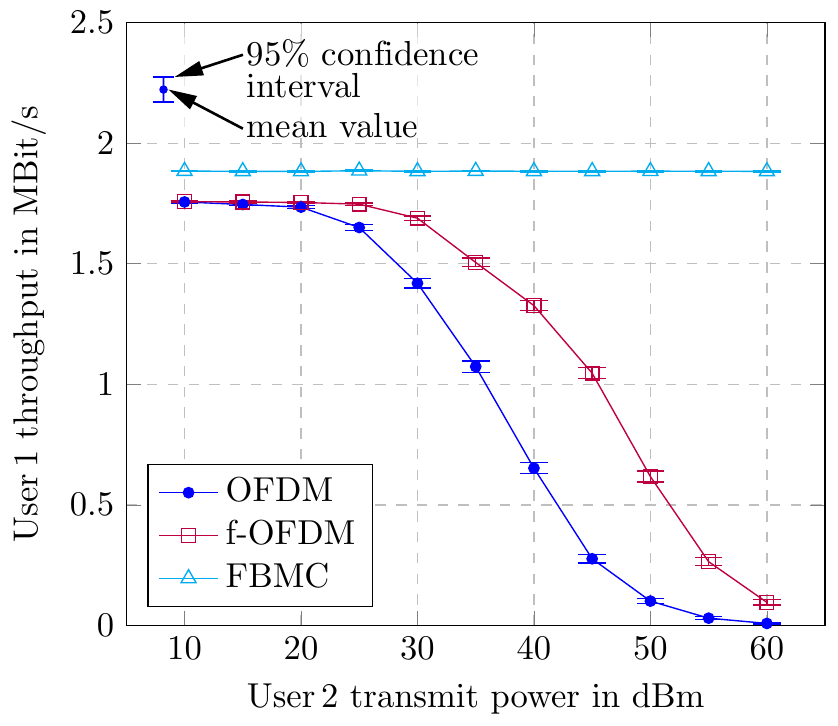}
	\caption{Simulation results of the multi-link scenario. Throughput of User\,1 is given for a sweep of transmit power of User\,2. The transmit power of User\,1 is $30\,$dBm.}
	\label{fig:inter_user_interference_throughput}
\end{figure}

We simulate User\,1's throughput for different waveforms, namely \ac{OFDM}, \ac{FOFDM} and \ac{FBMC} with simulation parameters summarized in~\cref{tab:inter_user_interference_parameters}.
User\,1's transmit power is $30\,$dBm and its path loss is chosen such that it has a high \ac{SNR} of approximately $40\,$dB.
Results for a sweep over the interfering User\,2's transmit power are shown in~\cref{fig:inter_user_interference_throughput}.
We observe that \ac{OFDM} leads to the highest impact of interference as it has the highest \ac{OOB} emissions of the three compared waveforms.
When \ac{OFDM} is employed, the throughput of User\,1 is already decreasing significantly for a transmit power of $30\,$dBm of the interfering User\,2.
For \ac{FOFDM}, the impact of interference is also severe, however, due to the filtering and the reduced \ac{OOB} emissions the drop in throughput occurs only at higher transmit powers of User\,2 compared to \ac{OFDM}.
If both users utilize \ac{FBMC}, then \ac{OOB} emissions decrease rapidly such that the $60\,$kHz guard band is sufficient to mitigate \ac{IUI}.
Further, we observe that \ac{FBMC} leads to a higher spectral efficiency compared to \ac{OFDM} and \ac{FOFDM} as it deploys no \ac{CP}.
The transmit power values of User\,2 are swept up to very high values in this simulation.
Please note, that this models a situation where the interfering user is close to the \ac{BS}.

\begin{table}[ht!]
	\caption{Simulation parameters for the multi-link simulation scenario. The effects of \acs*{IUI} are investigated for three different waveforms.}
	\label{tab:inter_user_interference_parameters}
	\begin{tabular}{l|c|c|c}
		\hline \hline
		\textbf{parameter} 		& \multicolumn{3}{c}{\textbf{value}} \\ \hline
		waveform 				& \acs*{OFDM} 		& \acs*{FOFDM} 		& \acs*{FBMC} \\
		filter type/length 		& - 				&  7.14\,\textmu s		& \!PHYDYAS-OQAM\! \\
		CP length 				& 4.76\,\textmu s 	&  4.76\,\textmu s		& - \\
		\hline
		subcarrier spacing 		& \multicolumn{3}{c}{User\,1: 15\,kHz, User\,2: 30\,kHz} \\
		guard band 				& \multicolumn{3}{c}{2$\times$15\,kHz\,+\,1$\times$30\,kHz\,=\,60\,kHz} \\
		bandwidth per user\!	& \multicolumn{3}{c}{34$\times$15\,kHz\,=\,17$\times$30\,kHz\,=\,0.51\,MHz} \\ 
		modulation/coding 		& \multicolumn{3}{c}{64\,\acs*{QAM}/\acs*{LDPC}, r\,=\,0.65 (\acs*{CQI}\,12)} \\
		channel model 			& \multicolumn{3}{c}{block fading Pedestrian A} \\
		\hline \hline
	\end{tabular}
\end{table}

\subsection{Non-Orthogonal Multiple Access}
Massive connectivity and low latency operation are one of the main drivers for future communications systems.
One promising solution addressing these requirements is \ac{NOMA}~\cite{Ding2017}.
It allows multiple \acp{UE} to share the same orthogonal resources in a non-orthogonal manner.
This increases the number of concurrent \acp{UE} and allows them to transmit more often.
Currently, the simulator supports a downlink version based on the \ac{3GPP} \ac{MUST} item~\cite{3GPP_TR_36859}, with more schemes planned for future releases of the simulator.
\Ac{MUST} allows the \ac{BS} to transmit to two \acp{UE} using the same frequency, time, and space by superimposing them in the power-domain.
One of those \acp{UE} has good channel conditions (Near\ac{UE}), while the other one has bad channel conditions (Far\ac{UE}), such as a cell-edge \ac{UE}.
The standard defines three power-ratios that control how much power is allocated to each \ac{UE}.
In either case, the Far\ac{UE} gets most of the power in order to help it overcome its harsh conditions.
At the receiver side, the interference from the high power \ac{UE} is mitigated by means of \ac{SIC}, or by directly applying \ac{ML} detection on the superimposed composite constellation.

\begin{figure}
	\centering
	\includegraphics[width=76mm]{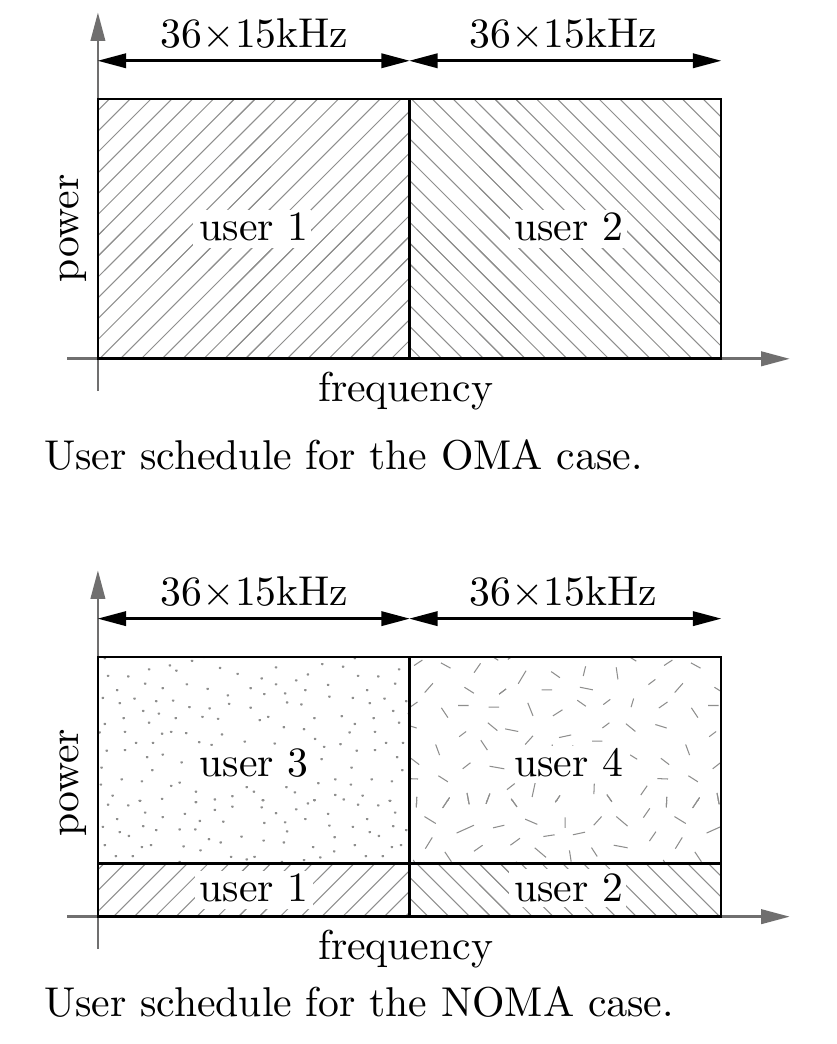}
	\caption{User assignment for the \acs*{NOMA} simulation scenario.}
	\label{fig:noma_scenarioSchedule}
\end{figure}

The remainder of this section is dedicated to an example scenario, introducing the concept of \ac{NOMA} support within our simulator.
In order to demonstrate the gain provided by the \ac{3GPP} \ac{MUST} scheme, we set up the following scenario: two cells, one operating with \ac{OMA}, and the other one with \ac{NOMA} based on \ac{MUST}.
In each cell, the \ac{BS} splits the bandwidth equally across two strong \acp{UE}; however, since the second \ac{BS} supports \ac{MUST}, it can superimpose those two strong \acp{UE} with additional two weak (cell-edge) \acp{UE}, thus providing a cell overloading of \(200\%\).~\cref{fig:noma_scenarioSchedule} illustrates the scheduling of the users in the two cells. Notice how the two additional users in the \ac{NOMA} case occupy the same resources as the main ones, but have a much higher allocated power. The notion of strong and weak \acp{UE} is achieved by choosing an appropriate path-loss for each \ac{UE}'s link.
\cref{tab:nomaExampleTable} summarizes the simulation parameters for this scenario.

\begin{table}[ht!]
	\caption{The simulation parameters for the \ac{NOMA} example scenario.}
	\label{tab:nomaExampleTable}
	\begin{tabular}{l|c|>{\centering\arraybackslash}p{30mm}}
		\hline \hline
		\textbf{parameter} 		& \multicolumn{2}{c}{\textbf{value}} \\ \hline
		cells				    & \acs*{OMA} 		& \acs*{NOMA}   \\
		number of users 		& 2 				&  4 (2 strong, 2 cell-edge)	\\
		path-loss				& 80, 90\,dB		&  strong: 80, 90\,dB \newline cell-edge: 110, 115\,dB	 \\
		\ac{NOMA} receiver		& -					&  \acs*{ML}	 \\
		\ac{MUST} power-ratio	& -					&  fixed (second ratio)	 \\ \hline
		bandwidth 				& \multicolumn{2}{c}{1.4\,MHz (72 subcarriers)} \\
		waveform/coding 		& \multicolumn{2}{c}{\ac{OFDM}, \acs*{LDPC}} \\
		\ac{MIMO} mode 			& \multicolumn{2}{c}{2$\times$2 \acs*{CLSM}} \\
		modulation/code rate    & \multicolumn{2}{c}{adaptive (\acs*{CQI} based)} \\ 
		feedback delay          & \multicolumn{2}{c}{no delay (ideal)} \\
		channel model 			& \multicolumn{2}{c}{Pedestrian A} \\
		\hline \hline
	\end{tabular}
\end{table}

\begin{figure}
	\centering
	\includegraphics[width=76mm]{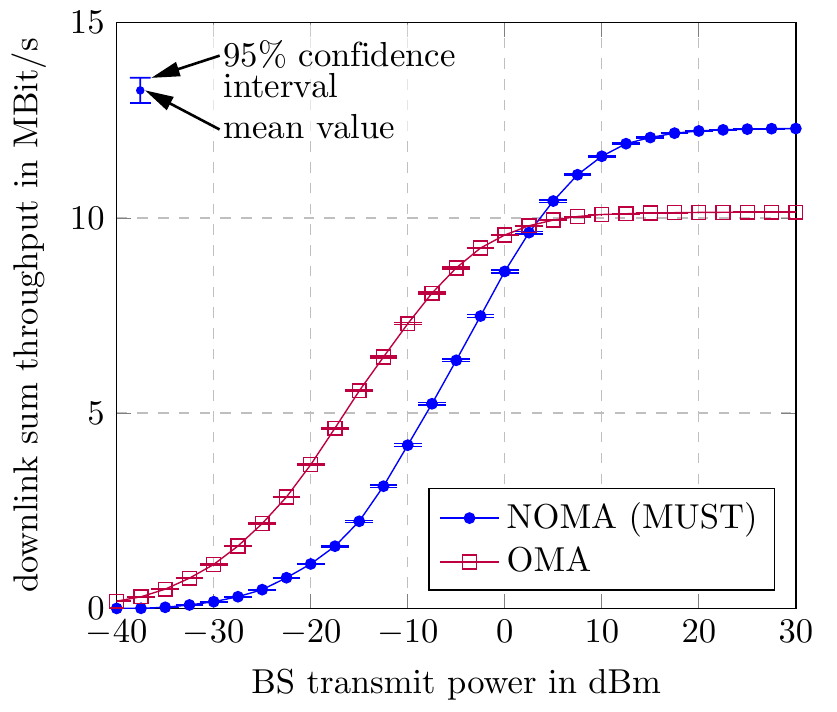}
	\caption{Simulation results of the \acs*{NOMA} simulation example. We show the downlink sum-throughput of the \ac{OMA} and \ac{NOMA} cells versus the transmit power of the \ac{BS}s.}
	\label{fig:must_sumthroughput}
\end{figure}

In~\cref{fig:must_sumthroughput}, the downlink sum-throughput for both the \ac{OMA} and \ac{NOMA} cells versus the transmit power of the \ac{BS}s is plotted.
When the transmit power of the \ac{BS} is low, we observe that \ac{MUST} is not beneficial, as the interference between the superimposed \acp{UE} has a substantial impact on the performance.
However, once the transmit power is sufficiently high, the receiver can carry out the interference suppression more effectively, leading to a considerable gain in the throughput.
An improvement of approximately \(20\%\) is observed at the transmit power of \(15\)\,dBm.
This corresponds to the \ac{SNR}s of \(51.6\) dB and \(41.6\)\,dB for the strong \acp{UE}, and \(21.6\) dB and \(16.6\)\,dB for the weak ones.
The \ac{SNR} here is with respect to the total superimposed received signal.
We conclude that, in general, MUST allows the \ac{BS} to support more \acp{UE} in the downlink, and this, combined with a sufficiently high transmit power, leads to an improved spectral efficiency.

\section{Conclusion} \label{sec:conclusion}
For the evolution of mobile communications from \ac{LTE} to \ac{5G} and beyond, an \ac{LL} simulation is an important tool, enabling research and development of advanced \ac{PHY} methods.
In this contribution, we introduce the \ac{VLLS}, which is freely available for researchers to support research and enhance reproducibility.
We give a general overview of the \ac{VLLS}, while further supporting documents and details are available at~\cite{vccsurl}.
Further, to outline the overall functionality, examples for specific features are provided, which are included in the simulators download package for increased reproducibility.
Our simulator supports standard compliant simulation scenarios and parameter settings, according to current communication specifications such as \ac{LTE} or \ac{5G} \ac{NR}.
Further, versatile functionality in terms of \ac{PHY} procedures and methods also allows simulation and investigation of potential candidate \ac{PHY} schemes beyond \ac{5G}.
The flexible and modular implementation additionally allows for easy augmentation of the simulator, e.g., implementation of additional features, making it a valuable tool for mobile communications research.



\begin{backmatter}

%

\section*{Acknowledgements}
This work has been funded by the Christian Doppler Laboratory for Dependable Wireless Connectivity for the Society in Motion.
The financial support by the Austrian Federal Ministry of Science, Research and Economy and the National Foundation for Research, Technology and Development is gratefully acknowledged.




\newcommand{\BMCxmlcomment}[1]{}

\BMCxmlcomment{

<refgrp>

<bibl id="B1">
  <title><p>The {V}ienna {LTE} Simulators --- {E}nabling {R}eproducibility in
  {W}ireless {C}ommunications {R}esearch</p></title>
  <aug>
    <au><snm>Mehlf\"uhrer</snm><fnm>C.</fnm></au>
    <au><snm>Ikuno</snm><fnm>J. C.</fnm></au>
    <au><snm>Simko</snm><fnm>M.</fnm></au>
    <au><snm>Schwarz</snm><fnm>S.</fnm></au>
    <au><snm>Rupp</snm><fnm>M.</fnm></au>
  </aug>
  <source>{EURASIP} {J}ournal on {A}dvances in {S}ignal {P}rocessing ({JASP})
  special issue on Reproducible Research</source>
  <pubdate>2011</pubdate>
  <volume>2011</volume>
  <issue>1</issue>
  <fpage>1</fpage>
  <lpage>-14</lpage>
</bibl>

<bibl id="B2">
  <title><p>Vienna Cellular Communications Simulators</p></title>
  <aug>
    <au><cnm>{Institute of Telecommunications, TU Wien}</cnm></au>
  </aug>
  <url>www.tc.tuwien.ac.at/vccs/</url>
</bibl>

<bibl id="B3">
  <title><p>The Vienna LTE-Advanced Simulators: Up and Downlink, Link and
  System Level Simulation</p></title>
  <aug>
    <au><snm>Rupp</snm><fnm>M</fnm></au>
    <au><snm>Schwarz</snm><fnm>S</fnm></au>
    <au><snm>Taranetz</snm><fnm>M</fnm></au>
  </aug>
  <publisher>Singapore: Springer</publisher>
  <edition>1</edition>
  <series><title><p>Signals and Communication Technology</p></title></series>
  <pubdate>2016</pubdate>
</bibl>

<bibl id="B4">
  <title><p>System level simulation of {LTE} networks</p></title>
  <aug>
    <au><snm>Ikuno</snm><fnm>JC</fnm></au>
    <au><snm>Wrulich</snm><fnm>M</fnm></au>
    <au><snm>Rupp</snm><fnm>M</fnm></au>
  </aug>
  <source>IEEE Vehicular Technology Conference (VTC Spring)</source>
  <publisher>Taipei, Taiwan</publisher>
  <pubdate>2010</pubdate>
</bibl>

<bibl id="B5">
  <title><p>Runtime Precoding: Enabling Multipoint Transmission in
  {LTE}-Advanced System Level Simulations</p></title>
  <aug>
    <au><snm>Taranetz</snm><fnm>M</fnm></au>
    <au><snm>Blazek</snm><fnm>T</fnm></au>
    <au><snm>Kropfreiter</snm><fnm>T</fnm></au>
    <au><snm>M\"uller</snm><fnm>MK</fnm></au>
    <au><snm>Schwarz</snm><fnm>S</fnm></au>
    <au><snm>Rupp</snm><fnm>M</fnm></au>
  </aug>
  <source>IEEE Access</source>
  <pubdate>2015</pubdate>
  <volume>3</volume>
  <fpage>725</fpage>
  <lpage>-736</lpage>
</bibl>

<bibl id="B6">
  <title><p>Exploring the physical layer frontiers of cellular
  uplink</p></title>
  <aug>
    <au><snm>Z{\"o}chmann</snm><fnm>E</fnm></au>
    <au><snm>Schwarz</snm><fnm>S</fnm></au>
    <au><snm>Pratschner</snm><fnm>S</fnm></au>
    <au><snm>Nagel</snm><fnm>L</fnm></au>
    <au><snm>Lerch</snm><fnm>M</fnm></au>
    <au><snm>Rupp</snm><fnm>M</fnm></au>
  </aug>
  <source>EURASIP Journal on Wireless Communications and Networking</source>
  <pubdate>2016</pubdate>
  <volume>2016</volume>
  <issue>1</issue>
  <fpage>1</fpage>
  <lpage>-18</lpage>
  <url>http://dx.doi.org/10.1186/s13638-016-0609-1</url>
</bibl>

<bibl id="B7">
  <title><p>Pushing the Limits of {LTE}: A Survey on Research Enhancing the
  Standard</p></title>
  <aug>
    <au><snm>Schwarz</snm><fnm>S.</fnm></au>
    <au><snm>Ikuno</snm><fnm>JC</fnm></au>
    <au><snm>Simko</snm><fnm>M.</fnm></au>
    <au><snm>Taranetz</snm><fnm>M.</fnm></au>
    <au><snm>Wang</snm><fnm>Q.</fnm></au>
    <au><snm>Rupp</snm><fnm>M.</fnm></au>
  </aug>
  <source>{IEEE} Access</source>
  <pubdate>2013</pubdate>
  <volume>1</volume>
  <fpage>51</fpage>
  <lpage>62</lpage>
  <url>http://dx.doi.org/10.1109/ACCESS.2013.2260371</url>
</bibl>

<bibl id="B8">
  <title><p>An Overview of Massive {MIMO}: Benefits and Challenges</p></title>
  <aug>
    <au><snm>Lu</snm><fnm>L</fnm></au>
    <au><snm>Li</snm><fnm>G.Y.</fnm></au>
    <au><snm>Swindlehurst</snm><fnm>A.L.</fnm></au>
    <au><snm>Ashikhmin</snm><fnm>A.</fnm></au>
    <au><snm>Zhang</snm><fnm>R</fnm></au>
  </aug>
  <source>IEEE Journal of Selected Topics in Signal Processing</source>
  <pubdate>2014</pubdate>
  <volume>8</volume>
  <issue>5</issue>
  <fpage>742</fpage>
  <lpage>758</lpage>
</bibl>

<bibl id="B9">
  <title><p>Massive {MIMO} for next generation wireless systems</p></title>
  <aug>
    <au><snm>Larsson</snm><fnm>E.</fnm></au>
    <au><snm>Edfors</snm><fnm>O.</fnm></au>
    <au><snm>Tufvesson</snm><fnm>F.</fnm></au>
    <au><snm>Marzetta</snm><fnm>T.</fnm></au>
  </aug>
  <source>IEEE Communications Magazine</source>
  <pubdate>2014</pubdate>
  <volume>52</volume>
  <issue>2</issue>
  <fpage>186</fpage>
  <lpage>195</lpage>
</bibl>

<bibl id="B10">
  <title><p>Overview of full-dimension {MIMO} in {LTE-A}dvanced pro</p></title>
  <aug>
    <au><snm>Ji</snm><fnm>H</fnm></au>
    <au><snm>Kim</snm><fnm>Y</fnm></au>
    <au><snm>Lee</snm><fnm>J</fnm></au>
    <au><snm>Onggosanusi</snm><fnm>E</fnm></au>
    <au><snm>Nam</snm><fnm>Y</fnm></au>
    <au><snm>Zhang</snm><fnm>J</fnm></au>
    <au><snm>Lee</snm><fnm>B</fnm></au>
    <au><snm>Shim</snm><fnm>B</fnm></au>
  </aug>
  <source>IEEE Communications Magazine</source>
  <publisher>IEEE</publisher>
  <pubdate>2017</pubdate>
  <volume>55</volume>
  <issue>2</issue>
  <fpage>176</fpage>
  <lpage>-184</lpage>
</bibl>

<bibl id="B11">
  <title><p>Waveform and Numerology to Support {5G} Services and
  Requirements</p></title>
  <aug>
    <au><snm>Zaidi</snm><fnm>A. A.</fnm></au>
    <au><snm>Baldemair</snm><fnm>R.</fnm></au>
    <au><snm>Tullberg</snm><fnm>H.</fnm></au>
    <au><snm>Bjorkegren</snm><fnm>H.</fnm></au>
    <au><snm>Sundstrom</snm><fnm>L.</fnm></au>
    <au><snm>Medbo</snm><fnm>J.</fnm></au>
    <au><snm>Kilinc</snm><fnm>C.</fnm></au>
    <au><snm>Silva</snm><fnm>ID</fnm></au>
  </aug>
  <source>{IEEE} Communications Magazine</source>
  <pubdate>2016</pubdate>
  <volume>54</volume>
  <issue>11</issue>
  <fpage>90</fpage>
  <lpage>98</lpage>
</bibl>

<bibl id="B12">
  <title><p>{5G} Field Trials: {OFDM}-Based Waveforms and Mixed
  Numerologies</p></title>
  <aug>
    <au><snm>Guan</snm><fnm>P.</fnm></au>
    <au><snm>Wu</snm><fnm>D.</fnm></au>
    <au><snm>Tian</snm><fnm>T.</fnm></au>
    <au><snm>Zhou</snm><fnm>J.</fnm></au>
    <au><snm>Zhang</snm><fnm>X.</fnm></au>
    <au><snm>Gu</snm><fnm>L.</fnm></au>
    <au><snm>Benjebbour</snm><fnm>A.</fnm></au>
    <au><snm>Iwabuchi</snm><fnm>M.</fnm></au>
    <au><snm>Kishiyama</snm><fnm>Y.</fnm></au>
  </aug>
  <source>{IEEE} Journal on Selected Areas in Communications</source>
  <pubdate>2017</pubdate>
  <volume>35</volume>
  <issue>6</issue>
  <fpage>1234</fpage>
  <lpage>1243</lpage>
</bibl>

<bibl id="B13">
  <title><p>Non-Orthogonal Multiple Access for Large-Scale {5G} Networks:
  Interference Aware Design</p></title>
  <aug>
    <au><snm>Ali</snm><fnm>K. S.</fnm></au>
    <au><snm>Elsawy</snm><fnm>H.</fnm></au>
    <au><snm>Chaaban</snm><fnm>A.</fnm></au>
    <au><snm>Alouini</snm><fnm>M. S.</fnm></au>
  </aug>
  <source>{IEEE} Access</source>
  <pubdate>2017</pubdate>
  <volume>5</volume>
  <fpage>21204</fpage>
  <lpage>21216</lpage>
</bibl>

<bibl id="B14">
  <title><p>A Survey on Non-Orthogonal Multiple Access for {5G} Networks:
  Research Challenges and Future Trends</p></title>
  <aug>
    <au><snm>Ding</snm><fnm>Z.</fnm></au>
    <au><snm>Lei</snm><fnm>X.</fnm></au>
    <au><snm>Karagiannidis</snm><fnm>G. K.</fnm></au>
    <au><snm>Schober</snm><fnm>R.</fnm></au>
    <au><snm>Yuan</snm><fnm>J.</fnm></au>
    <au><snm>Bhargava</snm><fnm>V. K.</fnm></au>
  </aug>
  <source>{IEEE} Journal on Selected Areas in Communications</source>
  <pubdate>2017</pubdate>
  <volume>35</volume>
  <issue>10</issue>
  <fpage>2181</fpage>
  <lpage>2195</lpage>
</bibl>

<bibl id="B15">
  <title><p>An Overview of Signal Processing Techniques for Millimeter Wave
  {MIMO} Systems</p></title>
  <aug>
    <au><snm>Heath</snm><fnm>R. W.</fnm></au>
    <au><snm>González Prelcic</snm><fnm>N.</fnm></au>
    <au><snm>Rangan</snm><fnm>S.</fnm></au>
    <au><snm>Roh</snm><fnm>W.</fnm></au>
    <au><snm>Sayeed</snm><fnm>A. M.</fnm></au>
  </aug>
  <source>{IEEE} Journal of Selected Topics in Signal Processing</source>
  <pubdate>2016</pubdate>
  <volume>10</volume>
  <issue>3</issue>
  <fpage>436</fpage>
  <lpage>453</lpage>
</bibl>

<bibl id="B16">
  <title><p>Millimeter-wave beamforming as an enabling technology for {5G}
  cellular communications: theoretical feasibility and prototype
  results</p></title>
  <aug>
    <au><snm>Roh</snm><fnm>W.</fnm></au>
    <au><snm>Seol</snm><fnm>J. Y.</fnm></au>
    <au><snm>Park</snm><fnm>J.</fnm></au>
    <au><snm>Lee</snm><fnm>B.</fnm></au>
    <au><snm>Lee</snm><fnm>J.</fnm></au>
    <au><snm>Kim</snm><fnm>Y.</fnm></au>
    <au><snm>Cho</snm><fnm>J.</fnm></au>
    <au><snm>Cheun</snm><fnm>K.</fnm></au>
    <au><snm>Aryanfar</snm><fnm>F.</fnm></au>
  </aug>
  <source>IEEE Communications Magazine</source>
  <pubdate>2014</pubdate>
  <volume>52</volume>
  <issue>2</issue>
  <fpage>106</fpage>
  <lpage>113</lpage>
</bibl>

<bibl id="B17">
  <title><p>Exploring Coordinated Multipoint Beamforming Strategies for {5G}
  Cellular</p></title>
  <aug>
    <au><snm>Schwarz</snm><fnm>S</fnm></au>
    <au><snm>Rupp</snm><fnm>M</fnm></au>
  </aug>
  <source>IEEE Access</source>
  <pubdate>2014</pubdate>
  <volume>2</volume>
  <fpage>930</fpage>
  <lpage>946</lpage>
  <url>http://ieeexplore.ieee.org/stamp/stamp.jsp?tp=&arnumber=6888496</url>
</bibl>

<bibl id="B18">
  <title><p>Rate Splitting for {MIMO} Wireless Networks: A Promising
  {PHY}-Layer Strategy for {LTE} Evolution</p></title>
  <aug>
    <au><snm>Clerckx</snm><fnm>B</fnm></au>
    <au><snm>Joudeh</snm><fnm>H</fnm></au>
    <au><snm>Hao</snm><fnm>C</fnm></au>
    <au><snm>Dai</snm><fnm>M</fnm></au>
    <au><snm>Rassouli</snm><fnm>B</fnm></au>
  </aug>
  <source>IEEE Communications Magazine</source>
  <pubdate>2016</pubdate>
  <volume>54</volume>
  <issue>5</issue>
  <fpage>98</fpage>
  <lpage>105</lpage>
</bibl>

<bibl id="B19">
  <title><p>Interference Alignment and Its Applications: A Survey, Research
  Issues, and Challenges</p></title>
  <aug>
    <au><snm>Zhao</snm><fnm>N.</fnm></au>
    <au><snm>Yu</snm><fnm>F. R.</fnm></au>
    <au><snm>Jin</snm><fnm>M.</fnm></au>
    <au><snm>Yan</snm><fnm>Q.</fnm></au>
    <au><snm>Leung</snm><fnm>V. C. M.</fnm></au>
  </aug>
  <source>{IEEE} Communications Surveys Tutorials</source>
  <pubdate>2016</pubdate>
  <volume>18</volume>
  <issue>3</issue>
  <fpage>1779</fpage>
  <lpage>1803</lpage>
</bibl>

<bibl id="B20">
  <title><p>Simulating {LTE} cellular systems: An open-source
  framework</p></title>
  <aug>
    <au><snm>Piro</snm><fnm>G</fnm></au>
    <au><snm>Grieco</snm><fnm>LA</fnm></au>
    <au><snm>Boggia</snm><fnm>G</fnm></au>
    <au><snm>Capozzi</snm><fnm>F</fnm></au>
    <au><snm>Camarda</snm><fnm>P</fnm></au>
  </aug>
  <source>IEEE Transactions on Vehicular Technology</source>
  <publisher>IEEE</publisher>
  <pubdate>2011</pubdate>
  <volume>60</volume>
  <issue>2</issue>
  <fpage>498</fpage>
  <lpage>-513</lpage>
</bibl>

<bibl id="B21">
  <title><p>open{WNS} - open Wireless Network Simulator</p></title>
  <aug>
    <au><snm>B{\"u}ltmann</snm><fnm>D</fnm></au>
    <au><snm>M{\"u}hleisen</snm><fnm>M</fnm></au>
    <au><snm>Klagges</snm><fnm>K</fnm></au>
    <au><snm>Schinnenburg</snm><fnm>M</fnm></au>
  </aug>
  <source>European Wireless Conference</source>
  <pubdate>2009</pubdate>
  <fpage>205</fpage>
  <lpage>-210</lpage>
</bibl>

<bibl id="B22">
  <title><p>The {GTEC} {5G} link-level simulator</p></title>
  <aug>
    <au><snm>Dom{\'\i}nguez Bola{\~n}o</snm><fnm>T</fnm></au>
    <au><snm>Rodr{\'\i}guez Pi{\~n}eiro</snm><fnm>J</fnm></au>
    <au><snm>Garc{\'\i}a Naya</snm><fnm>JA</fnm></au>
    <au><snm>Castedo</snm><fnm>L</fnm></au>
  </aug>
  <source>International Workshop on Link-and System Level Simulations
  (IWSLS)</source>
  <pubdate>2016</pubdate>
  <fpage>1</fpage>
  <lpage>-6</lpage>
</bibl>

<bibl id="B23">
  <title><p>Network simulations with the ns-3 simulator</p></title>
  <aug>
    <au><snm>Henderson</snm><fnm>TR</fnm></au>
    <au><snm>Lacage</snm><fnm>M</fnm></au>
    <au><snm>Riley</snm><fnm>GF</fnm></au>
    <au><snm>Dowell</snm><fnm>C</fnm></au>
    <au><snm>Kopena</snm><fnm>J</fnm></au>
  </aug>
  <source>SIGCOMM demonstration</source>
  <pubdate>2008</pubdate>
  <volume>14</volume>
  <issue>14</issue>
  <fpage>527</fpage>
</bibl>

<bibl id="B24">
  <title><p>An {LTE} module for the ns-3 network simulator</p></title>
  <aug>
    <au><snm>Piro</snm><fnm>G</fnm></au>
    <au><snm>Baldo</snm><fnm>N</fnm></au>
    <au><snm>Miozzo</snm><fnm>M</fnm></au>
  </aug>
  <source>Proceedings of the 4th International ICST Conference on Simulation
  Tools and Techniques</source>
  <pubdate>2011</pubdate>
  <fpage>415</fpage>
  <lpage>-422</lpage>
</bibl>

<bibl id="B25">
  <title><p>{5G} mmWave module for the ns-3 network simulator</p></title>
  <aug>
    <au><snm>Mezzavilla</snm><fnm>M</fnm></au>
    <au><snm>Dutta</snm><fnm>S</fnm></au>
    <au><snm>Zhang</snm><fnm>M</fnm></au>
    <au><snm>Akdeniz</snm><fnm>MR</fnm></au>
    <au><snm>Rangan</snm><fnm>S</fnm></au>
  </aug>
  <source>Proceedings of the 18th ACM International Conference on Modeling,
  Analysis and Simulation of Wireless and Mobile Systems</source>
  <pubdate>2015</pubdate>
  <fpage>283</fpage>
  <lpage>-290</lpage>
</bibl>

<bibl id="B26">
  <title><p>End-to-End Simulation of {5G} {mmWave} Networks</p></title>
  <aug>
    <au><snm>Mezzavilla</snm><fnm>M</fnm></au>
    <au><snm>Zhang</snm><fnm>M</fnm></au>
    <au><snm>Polese</snm><fnm>M</fnm></au>
    <au><snm>Ford</snm><fnm>R</fnm></au>
    <au><snm>Dutta</snm><fnm>S</fnm></au>
    <au><snm>Rangan</snm><fnm>S</fnm></au>
    <au><snm>Zorzi</snm><fnm>M</fnm></au>
  </aug>
  <source>arXiv preprint arXiv:1705.02882</source>
  <pubdate>2017</pubdate>
</bibl>

<bibl id="B27">
  <title><p>{OpenAirInterface}: A flexible platform for {5G}
  research</p></title>
  <aug>
    <au><snm>Nikaein</snm><fnm>N</fnm></au>
    <au><snm>Marina</snm><fnm>MK</fnm></au>
    <au><snm>Manickam</snm><fnm>S</fnm></au>
    <au><snm>Dawson</snm><fnm>A</fnm></au>
    <au><snm>Knopp</snm><fnm>R</fnm></au>
    <au><snm>Bonnet</snm><fnm>C</fnm></au>
  </aug>
  <source>ACM SIGCOMM Computer Communication Review</source>
  <publisher>ACM</publisher>
  <pubdate>2014</pubdate>
  <volume>44</volume>
  <issue>5</issue>
  <fpage>33</fpage>
  <lpage>-38</lpage>
</bibl>

<bibl id="B28">
  <title><p>{QuaDRiGa}: A {3-D} multi-cell channel model with time evolution
  for enabling virtual field trials</p></title>
  <aug>
    <au><snm>Jaeckel</snm><fnm>S</fnm></au>
    <au><snm>Raschkowski</snm><fnm>L</fnm></au>
    <au><snm>B{\"o}rner</snm><fnm>K</fnm></au>
    <au><snm>Thiele</snm><fnm>L</fnm></au>
  </aug>
  <source>IEEE Transactions on Antennas and Propagation</source>
  <publisher>IEEE</publisher>
  <pubdate>2014</pubdate>
  <volume>62</volume>
  <issue>6</issue>
  <fpage>3242</fpage>
  <lpage>-3256</lpage>
</bibl>

<bibl id="B29">
  <title><p>{3-D} millimeter-wave statistical channel model for {5G} wireless
  system design</p></title>
  <aug>
    <au><snm>Samimi</snm><fnm>MK</fnm></au>
    <au><snm>Rappaport</snm><fnm>TS</fnm></au>
  </aug>
  <source>IEEE Transactions on Microwave Theory and Techniques</source>
  <publisher>IEEE</publisher>
  <pubdate>2016</pubdate>
  <volume>64</volume>
  <issue>7</issue>
  <fpage>2207</fpage>
  <lpage>-2225</lpage>
</bibl>

<bibl id="B30">
  <title><p>A novel millimeter-wave channel simulator and applications for {5G}
  wireless communications</p></title>
  <aug>
    <au><snm>Sun</snm><fnm>S</fnm></au>
    <au><snm>MacCartney</snm><fnm>GR</fnm></au>
    <au><snm>Rappaport</snm><fnm>TS</fnm></au>
  </aug>
  <source>International Conference on Communications (ICC)</source>
  <pubdate>2017</pubdate>
  <fpage>1</fpage>
  <lpage>-7</lpage>
</bibl>

<bibl id="B31">
  <title><p>{Evolved Universal Terrestrial Radio Access} ({E-UTRA}) physical
  channels and modulation</p></title>
  <aug>
    <au><cnm>Generation Partnership Project (3GPP)</cnm></au>
  </aug>
  <source>TS</source>
  <pubdate>2015</pubdate>
  <issue>36.211</issue>
</bibl>

<bibl id="B32">
  <title><p>{Technical Specification Group Radio Access Network; NR; Physical
  channels and modulation}</p></title>
  <aug>
    <au><cnm>Generation Partnership Project (3GPP)</cnm></au>
  </aug>
  <source>TS</source>
  <pubdate>2017</pubdate>
  <issue>38.211</issue>
</bibl>

<bibl id="B33">
  <title><p>{Evolved Universal Terrestrial Radio Access (E-UTRA); Multiplexing
  and channel coding}</p></title>
  <aug>
    <au><cnm>Generation Partnership Project (3GPP)</cnm></au>
  </aug>
  <source>TS</source>
  <pubdate>2017</pubdate>
  <issue>36.212</issue>
</bibl>

<bibl id="B34">
  <title><p>{Technical Specification Group Radio Access Network; NR;
  Multiplexing and channel coding}</p></title>
  <aug>
    <au><cnm>Generation Partnership Project (3GPP)</cnm></au>
  </aug>
  <source>TS</source>
  <pubdate>2017</pubdate>
  <issue>38.212</issue>
</bibl>

<bibl id="B35">
  <title><p>New construction and performance analysis of Polar codes over
  {AWGN} channels</p></title>
  <aug>
    <au><snm>Tahir</snm><fnm>B</fnm></au>
    <au><snm>Rupp</snm><fnm>M</fnm></au>
  </aug>
  <source>24th International Conference on Telecommunications (ICT)</source>
  <pubdate>2017</pubdate>
  <fpage>1</fpage>
  <lpage>4</lpage>
</bibl>

<bibl id="B36">
  <title><p>{Optimal decoding of linear codes for minimizing symbol error rate
  (Corresp.)}</p></title>
  <aug>
    <au><snm>Bahl</snm><fnm>L.</fnm></au>
    <au><snm>Cocke</snm><fnm>J.</fnm></au>
    <au><snm>Jelinek</snm><fnm>F.</fnm></au>
    <au><snm>Raviv</snm><fnm>J.</fnm></au>
  </aug>
  <source>IEEE Transactions on Information Theory</source>
  <pubdate>1974</pubdate>
  <volume>20</volume>
  <issue>2</issue>
  <fpage>284</fpage>
  <lpage>287</lpage>
</bibl>

<bibl id="B37">
  <title><p>{Optimum and sub-optimum detection of coded data disturbed by
  time-varying intersymbol interference [applicable to digital mobile radio
  receivers]}</p></title>
  <aug>
    <au><snm>Koch</snm><fnm>W.</fnm></au>
    <au><snm>Baier</snm><fnm>A.</fnm></au>
  </aug>
  <source>Global Telecommunications Conference (GLOBECOM)</source>
  <pubdate>1990</pubdate>
  <fpage>1679</fpage>
  <lpage>1684vol.3</lpage>
</bibl>

<bibl id="B38">
  <title><p>{Linearly approximated log-MAP algorithms for turbo
  decoding}</p></title>
  <aug>
    <au><snm>Cheng</snm><fnm>JF</fnm></au>
    <au><snm>Ottosson</snm><fnm>T.</fnm></au>
  </aug>
  <source>51st Vehicular Technology Conference Proceedings</source>
  <pubdate>2000</pubdate>
  <volume>3</volume>
  <fpage>2252</fpage>
  <lpage>2256vol.3</lpage>
</bibl>

<bibl id="B39">
  <title><p>{Good error-correcting codes based on very sparse
  matrices}</p></title>
  <aug>
    <au><snm>MacKay</snm><fnm>D. J. C.</fnm></au>
  </aug>
  <source>IEEE Transactions on Information Theory</source>
  <pubdate>1999</pubdate>
  <volume>45</volume>
  <issue>2</issue>
  <fpage>399</fpage>
  <lpage>431</lpage>
</bibl>

<bibl id="B40">
  <title><p>Reduced-Complexity Decoding of {LDPC} Codes</p></title>
  <aug>
    <au><snm>Chen</snm><fnm>J</fnm></au>
    <au><snm>Dholakia</snm><fnm>A.</fnm></au>
    <au><snm>Eleftheriou</snm><fnm>E.</fnm></au>
    <au><snm>Fossorier</snm><fnm>M. P. C.</fnm></au>
    <au><snm>Hu</snm><fnm>XY</fnm></au>
  </aug>
  <source>IEEE Transactions on Communications</source>
  <pubdate>2005</pubdate>
  <volume>53</volume>
  <issue>8</issue>
  <fpage>1288</fpage>
  <lpage>1299</lpage>
</bibl>

<bibl id="B41">
  <title><p>{High-throughput LDPC decoders}</p></title>
  <aug>
    <au><snm>Mansour</snm><fnm>M. M.</fnm></au>
    <au><snm>Shanbhag</snm><fnm>N. R.</fnm></au>
  </aug>
  <source>IEEE Transactions on Very Large Scale Integration (VLSI)
  Systems</source>
  <pubdate>2003</pubdate>
  <volume>11</volume>
  <issue>6</issue>
  <fpage>976</fpage>
  <lpage>996</lpage>
</bibl>

<bibl id="B42">
  <title><p>Optimized Message Passing Schedules for {LDPC} Decoding</p></title>
  <aug>
    <au><snm>Radosavljevic</snm><fnm>P.</fnm></au>
    <au><snm>Baynast</snm><fnm>A.</fnm></au>
    <au><snm>Cavallaro</snm><fnm>J. R.</fnm></au>
  </aug>
  <source>Conference Record of the Thirty-Ninth Asilomar Conference on Signals,
  Systems and Computers</source>
  <pubdate>2005</pubdate>
  <fpage>591</fpage>
  <lpage>595</lpage>
</bibl>

<bibl id="B43">
  <title><p>{Channel Polarization: A Method for Constructing Capacity-Achieving
  Codes for Symmetric Binary-Input Memoryless Channels}</p></title>
  <aug>
    <au><snm>Arikan</snm><fnm>E.</fnm></au>
  </aug>
  <source>IEEE Transactions on Information Theory</source>
  <pubdate>2009</pubdate>
  <volume>55</volume>
  <issue>7</issue>
  <fpage>3051</fpage>
  <lpage>3073</lpage>
</bibl>

<bibl id="B44">
  <title><p>List decoding of polar codes</p></title>
  <aug>
    <au><snm>Tal</snm><fnm>I.</fnm></au>
    <au><snm>Vardy</snm><fnm>A.</fnm></au>
  </aug>
  <source>IEEE International Symposium on Information Theory
  Proceedings</source>
  <pubdate>2011</pubdate>
  <fpage>1</fpage>
  <lpage>5</lpage>
</bibl>

<bibl id="B45">
  <title><p>{Technical Specification Group Radio Access Network; Study on New
  Radio ({NR}) access techology}</p></title>
  <aug>
    <au><cnm>Generation Partnership Project (3GPP)</cnm></au>
  </aug>
  <source>TR</source>
  <pubdate>2017</pubdate>
  <issue>38.912</issue>
</bibl>

<bibl id="B46">
  <title><p>Filter bank multicarrier modulation schemes for future mobile
  communications</p></title>
  <aug>
    <au><snm>Nissel</snm><fnm>R</fnm></au>
    <au><snm>Schwarz</snm><fnm>S</fnm></au>
    <au><snm>Rupp</snm><fnm>M</fnm></au>
  </aug>
  <source>IEEE Journal on Selected Areas in Communications</source>
  <publisher>IEEE</publisher>
  <pubdate>2017</pubdate>
  <volume>35</volume>
  <issue>8</issue>
  <fpage>1768</fpage>
  <lpage>-1782</lpage>
</bibl>

<bibl id="B47">
  <title><p>Subcarrier spacing-a neglected degree of freedom?</p></title>
  <aug>
    <au><snm>Schaich</snm><fnm>F</fnm></au>
    <au><snm>Wild</snm><fnm>T</fnm></au>
  </aug>
  <source>16th International Workshop on Signal Processing Advances in Wireless
  Communications (SPAWC)</source>
  <pubdate>2015</pubdate>
  <fpage>56</fpage>
  <lpage>-60</lpage>
</bibl>

<bibl id="B48">
  <title><p>{Evolved Universal Terrestrial Radio Access (E-UTRA); {LTE}
  physical layer; General description}</p></title>
  <aug>
    <au><cnm>Generation Partnership Project (3GPP)</cnm></au>
  </aug>
  <source>TS</source>
  <pubdate>2018</pubdate>
  <issue>36.201</issue>
</bibl>

<bibl id="B49">
  <title><p>{WOLA-OFDM}: a potential candidate for asynchronous
  {5G}</p></title>
  <aug>
    <au><snm>Zayani</snm><fnm>R</fnm></au>
    <au><snm>Medjahdi</snm><fnm>Y</fnm></au>
    <au><snm>Shaiek</snm><fnm>H</fnm></au>
    <au><snm>Roviras</snm><fnm>D</fnm></au>
  </aug>
  <source>IEEE Globecom Workshops (GC Wkshps)</source>
  <pubdate>2016</pubdate>
  <fpage>1</fpage>
  <lpage>-5</lpage>
</bibl>

<bibl id="B50">
  <title><p>Waveform contenders for {5G—OFDM} vs. {FBMC} vs.
  {UFMC}</p></title>
  <aug>
    <au><snm>Schaich</snm><fnm>F</fnm></au>
    <au><snm>Wild</snm><fnm>T</fnm></au>
  </aug>
  <source>6th International Symposium on Communications, Control and Signal
  Processing (ISCCSP)</source>
  <pubdate>2014</pubdate>
  <fpage>457</fpage>
  <lpage>-460</lpage>
</bibl>

<bibl id="B51">
  <title><p>Universal-filtered multi-carrier technique for wireless systems
  beyond {LTE}</p></title>
  <aug>
    <au><snm>Vakilian</snm><fnm>V</fnm></au>
    <au><snm>Wild</snm><fnm>T</fnm></au>
    <au><snm>Schaich</snm><fnm>F</fnm></au>
    <au><snm>Brink</snm><fnm>S</fnm></au>
    <au><snm>Frigon</snm><fnm>JF</fnm></au>
  </aug>
  <source>IEEE Globecom Workshops (GC Wkshps)</source>
  <pubdate>2013</pubdate>
  <fpage>223</fpage>
  <lpage>-228</lpage>
</bibl>

<bibl id="B52">
  <title><p>{UFMC} system performance analysis for discrete narrow-band private
  networks</p></title>
  <aug>
    <au><snm>Geng</snm><fnm>S</fnm></au>
    <au><snm>Xiong</snm><fnm>X</fnm></au>
    <au><snm>Cheng</snm><fnm>L</fnm></au>
    <au><snm>Zhao</snm><fnm>X</fnm></au>
    <au><snm>Huang</snm><fnm>B</fnm></au>
  </aug>
  <source>6th International Symposium on Microwave, Antenna, Propagation, and
  EMC Technologies (MAPE)</source>
  <pubdate>2015</pubdate>
  <fpage>303</fpage>
  <lpage>-307</lpage>
</bibl>

<bibl id="B53">
  <title><p>{OFDM} for {5G}: Cyclic prefix versus zero postfix, and filtering
  versus windowing</p></title>
  <aug>
    <au><snm>Venkatesan</snm><fnm>S</fnm></au>
    <au><snm>Valenzuela</snm><fnm>RA</fnm></au>
  </aug>
  <source>International Conference on Communications (ICC)</source>
  <pubdate>2016</pubdate>
  <fpage>1</fpage>
  <lpage>-5</lpage>
</bibl>

<bibl id="B54">
  <title><p>Filtered {OFDM}: A new waveform for future wireless
  systems</p></title>
  <aug>
    <au><snm>Abdoli</snm><fnm>J</fnm></au>
    <au><snm>Jia</snm><fnm>M</fnm></au>
    <au><snm>Ma</snm><fnm>J</fnm></au>
  </aug>
  <source>16th International Workshop on Signal Processing Advances in Wireless
  Communications {(SPAWC)}</source>
  <pubdate>2015</pubdate>
  <fpage>66</fpage>
  <lpage>-70</lpage>
</bibl>

<bibl id="B55">
  <title><p>{FBMC-OQAM} in Doubly-Selective Channels: A New Perspective on
  {MMSE} Equalization</p></title>
  <aug>
    <au><snm>Nissel</snm><fnm>R</fnm></au>
    <au><snm>Rupp</snm><fnm>M</fnm></au>
    <au><snm>Marsalek</snm><fnm>R</fnm></au>
  </aug>
  <source>IEEE International Workshop on Signal Processing Advances in Wireless
  Communications (SPAWC)</source>
  <publisher>Sapporo, Japan</publisher>
  <pubdate>2017</pubdate>
</bibl>

<bibl id="B56">
  <title><p>{MMSE} equalization for {FBMC} transmission over doubly-selective
  channels</p></title>
  <aug>
    <au><snm>Marijanovi{\'c}</snm><fnm>L</fnm></au>
    <au><snm>Schwarz</snm><fnm>S</fnm></au>
    <au><snm>Rupp</snm><fnm>M</fnm></au>
  </aug>
  <source>International Symposium on Wireless Communication Systems
  (ISWCS)</source>
  <pubdate>2016</pubdate>
  <fpage>170</fpage>
  <lpage>-174</lpage>
</bibl>

<bibl id="B57">
  <title><p>Block frequency spreading: A method for low-complexity {MIMO} in
  {FBMC-OQAM}</p></title>
  <aug>
    <au><snm>Nissel</snm><fnm>R</fnm></au>
    <au><snm>Blumenstein</snm><fnm>J</fnm></au>
    <au><snm>Rupp</snm><fnm>M</fnm></au>
  </aug>
  <source>IEEE International Workshop on Signal Processing Advances in Wireless
  Communications (SPAWC)</source>
  <publisher>Sapporo, Japan</publisher>
  <pubdate>2017</pubdate>
</bibl>

<bibl id="B58">
  <title><p>Calculation of the spatial preprocessing and link adaption feedback
  for {3rd Generation Partnership Project (3GPP)} {UMTS/LTE}</p></title>
  <aug>
    <au><snm>Schwarz</snm><fnm>S</fnm></au>
    <au><snm>Mehlf{\"u}hrer</snm><fnm>C</fnm></au>
    <au><snm>Rupp</snm><fnm>M</fnm></au>
  </aug>
  <source>6th conference on Wireless advanced (WiAD)</source>
  <pubdate>2010</pubdate>
  <fpage>1</fpage>
  <lpage>-6</lpage>
</bibl>

<bibl id="B59">
  <title><p>Mutual information based calculation of the precoding matrix
  indicator for {3rd Generation Partnership Project (3GPP)}
  {UMTS/LTE}</p></title>
  <aug>
    <au><snm>Schwarz</snm><fnm>S</fnm></au>
    <au><snm>Wrulich</snm><fnm>M</fnm></au>
    <au><snm>Rupp</snm><fnm>M</fnm></au>
  </aug>
  <source>International ITG Workshop on Smart Antennas (WSA)</source>
  <pubdate>2010</pubdate>
  <fpage>52</fpage>
  <lpage>-58</lpage>
</bibl>

<bibl id="B60">
  <title><p>{3GPP 3D MIMO} Channel Model: A Holistic Implementation Guideline
  for Open Source Simulation Tools</p></title>
  <aug>
    <au><snm>Ademaj</snm><fnm>F</fnm></au>
    <au><snm>Taranetz</snm><fnm>M</fnm></au>
    <au><snm>Rupp</snm><fnm>M</fnm></au>
  </aug>
  <source>EURASIP Journal on Wireless Communications and Networking</source>
  <pubdate>2016</pubdate>
  <volume>2016</volume>
  <issue>1</issue>
  <fpage>55</fpage>
  <url>http://dx.doi.org/10.1186/s13638-016-0549-9</url>
</bibl>

<bibl id="B61">
  <title><p>{Study on 3D channel model for LTE}</p></title>
  <aug>
    <au><cnm>Generation Partnership Project (3GPP)</cnm></au>
  </aug>
  <source>TR</source>
  <pubdate>2015</pubdate>
  <issue>36.873</issue>
</bibl>

<bibl id="B62">
  <title><p>{Technical Specification Group Radio Access Network; High Speed
  Downlink Packet Access: UE Radio Transmission and Reception}</p></title>
  <aug>
    <au><cnm>Generation Partnership Project (3GPP)</cnm></au>
  </aug>
  <source>TR</source>
  <pubdate>2002</pubdate>
  <issue>25.890</issue>
</bibl>

<bibl id="B63">
  <title><p>{Technical Specification Group Radio Access Network; Study on
  channel model for frequencies from 0.5 to 100GHz}</p></title>
  <aug>
    <au><cnm>Generation Partnership Project (3GPP)</cnm></au>
  </aug>
  <source>TR</source>
  <pubdate>2017</pubdate>
  <issue>38.901</issue>
</bibl>

<bibl id="B64">
  <title><p>{Technical Specification Group Radio Access Network; Evolved
  Universal Terrestrial Radio Access: Base Station radio transmission and
  reception}</p></title>
  <aug>
    <au><cnm>Generation Partnership Project (3GPP)</cnm></au>
  </aug>
  <source>TR</source>
  <pubdate>2017</pubdate>
  <issue>36.104</issue>
</bibl>

<bibl id="B65">
  <title><p>Simulation models with correct statistical properties for Rayleigh
  fading channels</p></title>
  <aug>
    <au><snm>Zheng</snm><fnm>YR</fnm></au>
    <au><snm>Xiao</snm><fnm>C</fnm></au>
  </aug>
  <source>IEEE Transactions on communications</source>
  <publisher>IEEE</publisher>
  <pubdate>2003</pubdate>
  <volume>51</volume>
  <issue>6</issue>
  <fpage>920</fpage>
  <lpage>-928</lpage>
</bibl>

<bibl id="B66">
  <title><p>Time-variant channel estimation using discrete prolate spheroidal
  sequences</p></title>
  <aug>
    <au><snm>Zemen</snm><fnm>T</fnm></au>
    <au><snm>Mecklenbr\"auker</snm><fnm>CF</fnm></au>
  </aug>
  <source>IEEE Transactions on signal processing</source>
  <publisher>IEEE</publisher>
  <pubdate>2005</pubdate>
  <volume>53</volume>
  <issue>9</issue>
  <fpage>3597</fpage>
  <lpage>-3607</lpage>
</bibl>

<bibl id="B67">
  <title><p>New analytical models and probability density functions for fading
  in wireless communications</p></title>
  <aug>
    <au><snm>Durgin</snm><fnm>GD</fnm></au>
    <au><snm>Rappaport</snm><fnm>TS</fnm></au>
    <au><snm>Wolf</snm><fnm>DAD</fnm></au>
  </aug>
  <source>IEEE Transactions on Communications</source>
  <publisher>IEEE</publisher>
  <pubdate>2002</pubdate>
  <volume>50</volume>
  <issue>6</issue>
  <fpage>1005</fpage>
  <lpage>-1015</lpage>
</bibl>

<bibl id="B68">
  <title><p>{Evolved Universal Terrestrial Radio Access (E-UTRA); User
  Equipment (UE) radio transmission and reception}</p></title>
  <aug>
    <au><cnm>Generation Partnership Project (3GPP)</cnm></au>
  </aug>
  <source>TS</source>
  <pubdate>2017</pubdate>
  <issue>36.101</issue>
</bibl>

<bibl id="B69">
  <title><p>Optimal Numerology in {OFDM} Systems Based on Imperfect Channel
  Knowledge</p></title>
  <aug>
    <au><snm>Marijanovi{\'c}</snm><fnm>L</fnm></au>
    <au><snm>Schwarz</snm><fnm>S</fnm></au>
    <au><snm>Rupp</snm><fnm>M</fnm></au>
  </aug>
  <source>87th Vehicular Technology Conference: VTC2018-Spring</source>
  <pubdate>2018</pubdate>
</bibl>

<bibl id="B70">
  <title><p>{Technical Specification Group Radio Access Network; Study on
  Downlink Multiuser Superposition Transmission (MUST) for LTE}</p></title>
  <aug>
    <au><cnm>Generation Partnership Project (3GPP)</cnm></au>
  </aug>
  <source>TR</source>
  <pubdate>2015</pubdate>
  <issue>36.859</issue>
</bibl>

</refgrp>
} 

\end{backmatter}
\end{document}